\documentstyle[amstex,amssymb,epsfig,12pt]{article}

\begin{document}

\bigskip 
\begin{titlepage}
\bigskip \begin{flushright}
\end{flushright}


\vspace{1cm}

\begin{center}
{\Large \bf {Abelian Higgs Hair for AdS-Schwarzschild Black Hole        }}\\
\end{center}
\vspace{2cm}
\begin{center}
 M.H. Dehghani\footnote{%
EMail: hossein@@avatar.uwaterloo.ca; On leave from Physics
Dept., College of Sciences, Shiraz University, Shiraz, Iran},  A.M. Ghezelbash
\footnote{
EMail: amasoud@@avatar.uwaterloo.ca} and R. B. Mann{%
\footnote{%
EMail: mann@@avatar.uwaterloo.ca}} \\
Department of Physics, University of Waterloo, \\
Waterloo, Ontario N2L 3G1, CANADA\\
\vspace{1cm}
\today\\
\end{center}

\begin{abstract}
We show that the Abelian Higgs field equations in the background of the four dimensional
AdS-Schwarzschild black hole  have a vortex line solution.
This solution, which has axial symmetry,  is a generalization of the AdS 
spacetime Nielsen-Olesen string.  By a numerical study of the field equations, we 
show that black hole could support the Abelian Higgs field as its Abelian hair. 
Also,  we conside the self gravity of the  Abelian Higgs field both in the pure 
AdS spacetime and AdS-Schwarzschild black hole background and show that the 
effect of  string as a black hole hair is to induce a deficit angle
in the AdS-Schwarzschild black hole.

\end{abstract}
\end{titlepage}
\onecolumn

\begin{center}
\bigskip 
\end{center}

\section{Introduction}

The classical no-hair conjecture first proposed by Ruffini and Wheeler \cite
{Ruf} states that after a given distribution of matter collapses to form a
black hole, the only long range information of a black hole is its
electromagnetic charge, mass and angular momentum. In certain special cases
the conjecture has been verified. For example a scalar field minimally
coupled to gravity in asymptotically flat or de Sitter spacetimes cannot
provide hair for the black hole \cite{Sud},\cite{Torii1}.

While it is tempting to extend the no-hair theorem claim to all forms of
matter, it is known that some long range Yang-Mills and/or quantum hair
could be painted on the black holes \cite{Eli}. Explicit calculations have
been carried out which verify the existence of a long range Nielsen-Olesen
vortex solution as a stable hair for a Schwarzchild black hole in four
dimensions \cite{Achu}, although it might be argued that this situation
falls outside the scope of the classical no-hair theorem due to the non
trivial topology of the string configuration. More recently it has been
shown that an asymptotically flat black hole could be pierced by several
infinitely thin cosmic strings in a polyhedral configuration \cite{Frolov}.
These is much current interest in extending these considerations to anti de
Sitter spacetime, mainly due to the efforts of \ Maldacena \cite{Mal} and
Witten \cite{W} concerning the relation of the some large ${\cal N}$ gauge
theories in AdS spacetime and conformal field theories.

\medskip

Insofar as the no-hair theorem is concerned it has been shown that there
exists a solution to the $SU(2)$ Einstein-Yang-Mills equations which
describes a stable Yang-Mills hairy black hole that is asymptotically AdS 
\cite{Eli}. More recently we have shown that the $U(1)$ Higgs field
equations have a vortex solution in four dimensional AdS spacetime \cite{Deh}%
. \ More recently it has been shown that in asymptotically AdS spacetime,
that a black hole can have scalar hair \cite{Torii2}.

Motivated by these considerations, in this article we investigate possible
solutions of the Abelian-Higgs field equations in a four dimensional
AdS-Schwarzschild black hole background. While an analytical solution to
these equations appears to be intractable, we confirm but by numerical
calculation that AdS-Schwarzscild black hole could support a long range
cosmic string as its stable hair. The generalization to the multi-string
configurations as a black hole hairs also could be done.

In section two, we solve the first-order Einstein equations in the pure AdS$%
_{4}$ spacetime in the presence of a vortex solution. In section three, we
solve numerically the Abelian-Higgs equations in the Ads-Schwarzschild
background for different values of the cosmological constant and string
winding numbers. In section four, by studying the behaviour of the string
energy-momentum tensor, we find the effect of the vortex self gravity on the
AdS-Schwarzschild background metric. We argue in section five, that
AdS-Schwarzschild black hole could support a multi-string configuration. We
give some closing remarks in the final section.

\section{Vortex Self Gravity on AdS$_{4}$}

We consider first the effect of \ the vortex on the AdS$_{4}$ spacetime.
This entails finding the solutions of the coupled Einstein-Abelian Higgs
differential equations in AdS$_{4}$. This is a formidable problem even for
flat spacetime, and no exact solutions have been found \ for the flat
spacetime yet.

However\thinspace \ some physical results can be obtained by making some
approximations. First, we assume that the thickness of the vortex is much
smaller that all the other relevant length scales. Second, we assume that
the gravitational effects of the string are weak enough so that the
linearized Einstein-Abelian Higgs differential equations are applicable.

For convenience, in this section we use the following form of the metric of
AdS$_{4}.$

\begin{equation}
ds^{2}=-\widetilde{A}(r,\theta )^{2}dt^{2}+\widetilde{B}(r,\theta )^{2}d\phi
^{2}+\widetilde{C}(r,\theta )(\frac{dr^{2}}{1+\frac{r^{2}}{l^{2}}}
+r^{2}d\theta ^{2})  \label{ABCmetric}
\end{equation}
This metric in spherical coordinates is suitable for generalizing the vortex
self gravity in the presence of the AdS-Schwarzschild black hole. In the
absence of the vortex, we must have $A_{0}(r,\theta )=\sqrt{1+\frac{r^{2}}{
l^{2}}},B_{0}(r,\theta )=r\sin \theta ,C_{0}(r,\theta )=1$ , yielding the
well known metric of pure AdS$_{4}$.

Employing the two assumptions concerning the thickness of the vortex core
and its weak gravitational field, we solve numerically the Einstein field
equations, 
\begin{equation}
G_{\mu \nu }-\frac{3}{l^{2}}g_{\mu \nu }=-8\pi G{\cal T}_{\mu \nu }
\label{Einstein}
\end{equation}
to the first order in $\varepsilon =8\pi G,$ where ${\cal T}_{\mu \nu }$ is
the energy-momentum tensor of the Abelian Higgs field in the AdS background.
To first order of approximation by taking $g_{\mu \nu }\simeq g_{\mu \nu
}^{(0)}+g_{\mu \nu }^{(1)}$ , where $g_{\mu \nu }^{(0)}$ is the usual AdS$_{4%
\text{ }}$ metric, $\ g_{\mu \nu }^{(1)}$ is the first order correction to
the metric and writing 
\begin{equation}
\begin{tabular}{l}
$\widetilde{A}(r,\theta )=A_{0}(r,\theta )(1+\varepsilon A(r,\theta ))$ \\ 
$\widetilde{B}(r,\theta )=B_{0}(r,\theta )(1+\varepsilon B(r,\theta ))$ \\ 
$\widetilde{C}(r,\theta )=C_{0}(r,\theta )(1+\varepsilon C(r,\theta ))$%
\end{tabular}
\label{ABCexpa}
\end{equation}
we obtain corrections to the three functions $A_{0}(r,\theta
),B_{0}(r,\theta )$ and $C_{0}(r,\theta )$ in (\ref{ABCexpa}). Hence in the
first approximation the equations (\ref{Einstein}) become

\begin{equation}
G_{\mu \nu }^{(1)}-\frac{3}{l^{2}}g_{\mu \nu }^{(1)}=-{\cal T}_{\mu \nu
}^{(0)}  \label{Eineq}
\end{equation}
where ${\cal T}_{\mu \nu }^{(0)}$ is the energy momentum tensor of string
field in AdS$_{4\text{ }}$background metric, and $G_{\mu \nu }^{(1)}$ is the
correction to the Einstein tensor due to $g_{\mu \nu }^{(1)}$.

The rescaled components of the energy momentum tensor of string in the
background of AdS$_{4},$ are given by, 
\begin{equation}
\begin{tabular}{l}
$T_{t}^{t(0)}(\rho )=-\frac{1}{2}(\frac{dX}{d\rho })^{2}(1+\frac{\rho ^{2}}{
l^{2}})-\frac{1}{2}\frac{1}{\rho ^{2}}(\frac{dP}{d\rho })^{2}(1+\frac{\rho
^{2}}{l^{2}})-\frac{1}{2}\frac{P^{2}X^{2}}{\rho ^{2}}-(X^{2}-1)^{2}$ \\ 
$T_{\varphi }^{\varphi (0)}(\rho )=-\frac{1}{2}(\frac{dX}{d\rho })^{2}(1+%
\frac{\rho ^{2}}{l^{2}})+\frac{1}{2}\frac{1}{\rho ^{2}}(\frac{dP}{d\rho }
)^{2}(1+\frac{\rho ^{2}}{l^{2}})+\frac{1}{2}\frac{P^{2}X^{2}}{\rho ^{2}}
-(X^{2}-1)^{2}$ \\ 
$(T_{r}^{r(0)}+T_{\theta }^{\theta (0)})(\rho )=-\frac{P^{2}X^{2}}{\rho ^{2}}
-2(X^{2}-1)^{2}$%
\end{tabular}
\label{stresssph}
\end{equation}
where $X$ and $P$ are the solutions of \ the string fields \cite{Deh} and $%
\rho =r\sin \theta $.

\begin{figure}[tbp]
\begin{center}
\epsfig{file=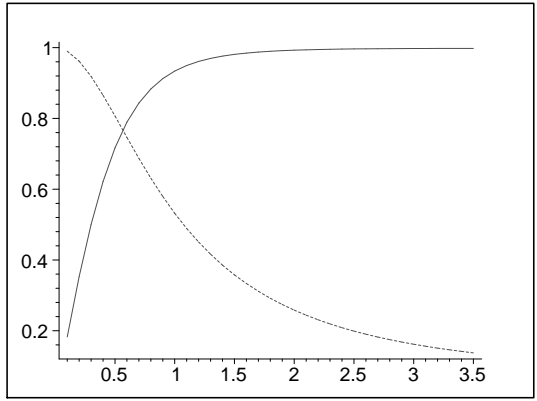,width=0.4\linewidth}
\end{center}
\caption{$X(\protect\rho )$ (solid) and $P(\protect\rho )$ (dotted) for $l=1$
}
\label{fig1}
\end{figure}

The Einstein equations (\ref{Eineq}) are 
\begin{equation}
\begin{tabular}{l}
$(1+\frac{\rho ^{2}}{l^{2}})\frac{d^{2}B}{d\rho ^{2}}+2\frac{dB}{d\rho }(%
\frac{1}{\rho }+\frac{2\rho }{l^{2}})+\frac{1}{2}(1+\frac{\rho ^{2}}{l^{2}})%
\frac{d^{2}C}{d\rho ^{2}}+\frac{\rho }{l^{2}}\frac{dC}{d\rho }-\frac{3C}{%
l^{2}}=T_{t}^{t(0)}$ \\ 
$(1+\frac{\rho ^{2}}{l^{2}})\frac{d^{2}A}{d\rho ^{2}}+\frac{4\rho }{l^{2}}%
\frac{dA}{d\rho }+\frac{1}{2}(1+\frac{\rho ^{2}}{l^{2}})\frac{d^{2}C}{d\rho
^{2}}+\frac{\rho }{l^{2}}\frac{dC}{d\rho }-\frac{3C}{l^{2}}=T_{\varphi
}^{\varphi (0)}$ \\ 
$(1+\frac{\rho ^{2}}{l^{2}})(\frac{d^{2}A}{d\rho ^{2}}+\frac{d^{2}B}{d\rho
^{2}})+\frac{2}{\rho }(\frac{dA}{d\rho }+\frac{dB}{d\rho })(1+3\frac{\rho
^{2}}{l^{2}})-\frac{6C}{l^{2}}=T_{r}^{r(0)}+T_{\theta }^{\theta (0)}$%
\end{tabular}
\label{eins}
\end{equation}
Solving equations of motions of the Abelian Higgs fields \cite{Deh}
numerically gives the following graphs for $X$ and $P$ fields (\ref{fig1}) 

\begin{figure}[tbp]
\begin{center}
\epsfig{file=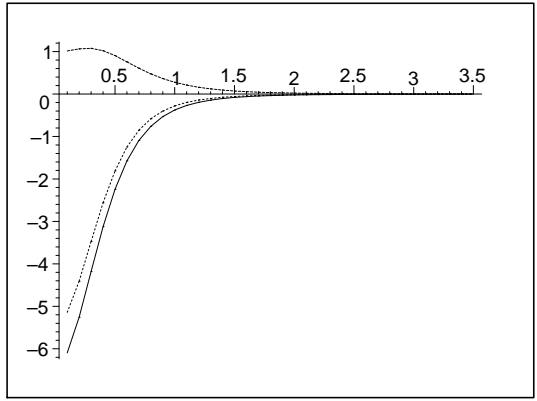,width=0.4\linewidth}
\end{center}
\caption{$T_{t}^{t(0)}$(solid),$T_{\protect\varphi }^{\protect\varphi (0)}$%
(dashed)$,T_{r}^{r(0)}+T_{\protect\theta }^{\protect\theta (0)}$
(dash-dotted) curves for $l=1$}
\label{fig2}
\end{figure}

By using the equations (\ref{stresssph}), one could obtain the diagram (\ref
{fig2}) showing the behaviour of stress tensor components.

\begin{figure}[tbp]
\begin{center}
\epsfig{file=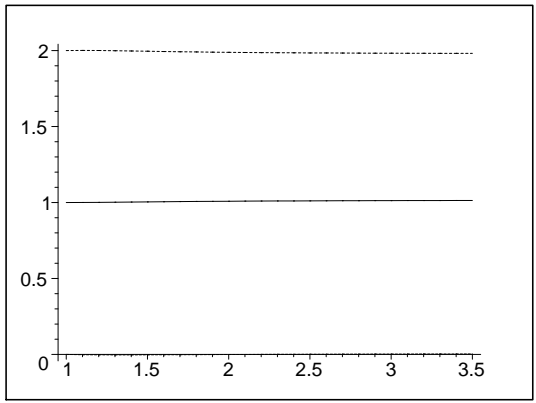,width=0.4\linewidth}
\end{center}
\caption{$A$(solid),$B$(dotted)$,C$ touches the horizontal axis}
\label{fig3}
\end{figure}

Then solving the coupled differential equations (\ref{eins}) gives the
behaviour of functions $A(\rho ),B(\rho )$ and $C(\rho )$ versus $\rho .$ (%
\ref{fig3})

Hence by a redefinition of time coordinate in (\ref{ABCmetric}) the metric
can be rewritten as 
\begin{equation}
ds^{2}=-(1+\frac{r^{2}}{l^{2}})dt^{2}+\frac{dr^{2}}{1+\frac{r^{2}}{l^{2}}}%
+r^{2}(d\theta ^{2}+\alpha ^{2}\sin ^{2}\theta d\phi ^{2})  \label{adsdef}
\end{equation}
which is the metric of AdS space with deficit angle.

\bigskip

\section{Abelian Higgs Vortex in AdS$-$Schwarzschild Black Hole}

We take the Abelian Higgs Lagrangian in AdS-Schwarzschild as follows,

\begin{equation}
{\cal L}(\Phi ,A_{\mu })=-\frac{1}{2}({\cal D}_{\mu }\Phi )^{\dagger }{\cal D%
}^{\mu }\Phi -\frac{1}{16\pi }{\cal F}_{\mu \nu }{\cal F}^{\mu \nu }-\xi
(\Phi ^{\dagger }\Phi -\eta ^{2})^{2}  \label{Lag}
\end{equation}
where $\Phi $ is a complex \ scalar Klein-Gordon field, ${\cal F}_{\mu \nu }$
is the field strength of \ the electromagnetic field $A_{\mu }$ and ${\cal D}%
_{\mu }=\nabla _{\mu }+ieA_{\mu }$ in which $\nabla _{\mu }$ is the
covariant derivative. We employ Planck units $G=\hbar =c=1$ which implies
that the Planck mass is equal to unity, and write the AdS-Schwarzschild
black hole metric in the form 
\begin{equation}
ds^{2}=-(1-\frac{2m}{r}+\frac{r^{2}}{l^{2}})dt^{2}+\frac{1}{(1-\frac{2m}{r}+%
\frac{r^{2}}{l^{2}})}dr^{2}+r^{2}(d\theta ^{2}+\sin ^{2}\theta \,d\varphi
^{2})  \label{adsmetr1}
\end{equation}
\qquad where $\ m$ is the black hole mass and the cosmological constant $%
\Lambda $ is equal to $\frac{-3}{l^{2}}.$ Defining the real fields $X(x^{\mu
}),\omega (x^{\mu }),P_{\mu }(x^{\nu })$ by the following equations

\begin{equation}
\begin{tabular}{l}
$\Phi (x^{\mu })=\eta X(x^{\mu })e^{i\omega (x^{\mu })}$ \\ 
$A_{\mu }(x^{\nu })=\frac{1}{e}(P_{\mu }(x^{\nu })-\nabla _{\mu }\omega
(x^{\mu }))$%
\end{tabular}
\label{XPomegadef}
\end{equation}
and employing a suitable choice of gauge, one could rewrite the Lagrangian ( 
\ref{Lag}) and the equations of motion in terms of these fields as: 
\begin{equation}
{\cal L(}X,P_{\mu })=-\frac{\eta ^{2}}{2}(\nabla _{\mu }X\,\nabla ^{\mu
}X+X^{2}P_{\mu }P^{\mu })-\frac{1}{16\pi e^{2}}F_{\mu \nu }F^{\mu \nu }-\xi
\eta ^{4}(X^{2}-1)^{2}  \label{Lag2}
\end{equation}

\begin{equation}
\begin{tabular}{l}
$\nabla _{\mu }\nabla ^{\mu }X-XP_{\mu }P^{\mu }-4\xi \eta ^{2}X(X^{2}-1)=0$
\\ 
$\nabla _{\mu }F^{\mu \nu }-4\pi e^{2}\eta ^{2}P^{\nu }X^{2}=0$%
\end{tabular}
\label{eqmo2}
\end{equation}
where $F^{\mu \nu }=\nabla ^{\mu }P^{\nu }-\nabla ^{\nu }P^{\mu }$ is the
field strength of the corresponding gauge field $P^{\mu }$. Note that the
real field $\omega $ is not itself a physical quantity. Superficially it
appears not to contain any physical information. However if $\omega $ is not
single valued this is no longer the case, and the resultant solutions are
referred to as vortex solutions \cite{NO}. \ In this case the requirement
that $\Phi $ field be single-valued implies that the line integral of $%
\omega $ over any closed loop is $\pm 2\pi n$ where $n$ is an integer. In
this case the flux of electromagnetic field $\Phi _{H\text{ \ }}$passing
through such a closed loop is quantized with quanta $2\pi /e.$

We seek a vortex solution for the Abelian Higgs Lagrangian (\ref{Lag2}) in
the background of AdS-Schwarzschild black hole.This solution can be
interpreted as a string piercing to the black hole (\ref{adsmetr1}).
Considering the static case of winding number $N$ with the gauge choice, 
\begin{equation}
P_{\mu }(r,\theta )=(0;0,0,NP(r,\theta ))  \label{Pgauge}
\end{equation}
and rescaling 
\begin{equation}
\varkappa \rightarrow \frac{\varkappa }{\sqrt{\xi }\eta }  \label{rescale}
\end{equation}
where $\varkappa =r,l,m,$ the equations of motion (\ref{eqmo2}) are

\begin{equation}
\begin{tabular}{l}
$(1-\frac{2m}{r}+\frac{r^{2}}{l^{2}})\frac{\partial ^{2}X(r,\theta )}{%
\partial r^{2}}+\frac{2}{r}(1-\frac{m}{r}+\frac{2r^{2}}{l^{2}})\frac{%
\partial X(r,\theta )}{\partial r}+\frac{1}{r^{2}}\frac{\partial
^{2}X(r,\theta )}{\partial \theta ^{2}}+\frac{1}{r^{2}}\frac{\partial
X(r,\theta )}{\partial \theta }\cot \theta -\frac{1}{2}(X^{3}(r,\theta )$ \\ 
$-X(r,\theta ))-N^{2}\frac{X(r,\theta )P^{2}(r,\theta )}{r^{2}\sin
^{2}\theta }=0$%
\end{tabular}
\label{eqx}
\end{equation}

\begin{equation}
\begin{tabular}{l}
$(1-\frac{2m}{r}+\frac{r^{2}}{l^{2}})\frac{\partial ^{2}P(r,\theta )}{
\partial r^{2}}+\frac{2}{r}(\frac{m}{r}+\frac{r^{2}}{l^{2}})\frac{\partial
P(r,\theta )}{\partial r}+\frac{1}{r^{2}}\frac{\partial ^{2}P(r,\theta )}{%
\partial \theta ^{2}}-\frac{\cot \theta }{r^{2}}\frac{\partial P(r,\theta )}{
\partial \theta }$ \\ 
$-\alpha P(r,\theta )X^{2}(r,\theta )=0$%
\end{tabular}
\label{eqp}
\end{equation}

In the above relation (\ref{eqp}), $\alpha =\frac{4\pi e^{2}}{\xi }.$ It
must be noted that even in the case of $m=0$, \ no exact analytic solutions
have been known for equations (\ref{eqp}) and (\ref{eqx}). So, in the rest
of this section, we seek the existence of vortex solutions for the above
coupled non linear partial differential equations. First, we consider the 
{\it thin }string with winding number one, in which one can assume $m>>1.$
The thicker vortex and larger winding number will be discussed later in this
section. Employing the ansatz

\begin{equation}
P(r,\theta )=P(\rho ),\text{ \ }X(r,\theta )=X(\rho )  \label{PXcylsym}
\end{equation}
where $\rho =r\sin \theta $,\ we get the following equations:

\begin{equation}
(1+\frac{\rho ^{2}}{l^{2}})\frac{d^{2}X}{d\rho ^{2}}+(\frac{1}{\rho }+\frac{
4\rho }{l^{2}})\frac{dX}{d\rho }-\frac{1}{2}X(X^{2}-1)-\frac{N^{2}}{\rho ^{2}%
}XP^{2}-2\frac{m\rho ^{2}}{r^{3}}(\frac{d^{2}X}{d\rho ^{2}}+\frac{1}{\rho }%
\frac{dX}{d\rho })=0  \label{eqx2}
\end{equation}

\begin{equation}
(1+\frac{\rho ^{2}}{l^{2}})\frac{d^{2}P}{d\rho ^{2}}+\frac{dP}{d\rho }(-%
\frac{1}{\rho }+\frac{2\rho }{l^{2}})-\alpha PX^{2}-2\frac{m\rho ^{2}}{r^{3}}
(\frac{d^{2}P}{d\rho ^{2}}-\frac{1}{\rho }\frac{dP}{d\rho })=0  \label{eqp2}
\end{equation}
As expected, in the limit $l\rightarrow \infty $ equations (\ref{eqx2}) and (%
\ref{eqp2}) reduce to the asymptotically flat case discussed in \cite{Achu}.
In this instance the vortex solutions of the Abelian Higgs equations in flat
spacetime (without a black hole) satisfy the $l\rightarrow \infty $ limit of
equations (\ref{eqp2} ) and (\ref{eqx2}) up to errors which are proportional
to the $\frac{m\rho ^{2}}{r^{3}}\approx \frac{m}{r^{3}}.$ These errors are
very tiny far from the black hole horizon, whereas near the horizon $%
r\approx r_{H}=2m,$ they are of the order of $\frac{1}{m^{2}},$ which is
negligible for large mass black holes. This observation suggested that a
string vortex solution could be painted to the horizon of a Schwarzschild
black hole. This conjecture has been further supported by numerical
calculations \cite{Achu} which show the existence of vortex solutions of the
Abelian Higgs equations in the background of the Schwarzschild black hole.
These calculations explicitly demonstrate that a cosmic string can pierce a
black hole for a variety of black hole masses and vortex winding numbers.

For\ finite $l$ ,we showed in our previous paper \cite{Deh} that the Abelian
Higgs equations of motion \ in the background of anti-de-Sitter spacetime ( (%
\ref{eqx2}) and (\ref{eqp2}) in the limit of \ $m=0$) have vortex solutions
(denoted by $X_{0}$ and $P_{0}$) with core radius $\rho \approx O(1)$ . The
functions $X_{0}$ and $P_{0}$ satisfy eqs. (\ref{eqx2}) and (\ref{eqp2}) up
to errors which are proportional to $\frac{m\rho ^{2}}{r^{3}}\approx \frac{m%
}{r^{3}}.$ These errors go to zero far from the black hole. However near the
horizon of \ a large mass black hole, $\ r\approx r_{H}\approx m^{1/3}$ ,
the term $\frac{m}{r^{3}}$ is at least of \ the order of unity, and so the
possibility of painting a string vortex solution to the horizon for finite $%
l $ remains unclear. Then we should go to perform the numerical calculations
to show the existance of vortex solution on, near and far from the horizon.

\subsection{Numerical Solutions}

We pay attention now to the numerical solutions of the equations (\ref{eqp})
and (\ref{eqx}) outside the black hole horizon. First, we must take
appropriate boundary conditions. At the large distance from the horizon, we
demand that our solutions go to the \ solutions of the vortex equations in
AdS spacetime given in \cite{Deh}. This means that we demand $X\rightarrow 1$
and $P\rightarrow 0$ as $\rho $ goes to infinity. On the symmetry axis of
the string and beyond the radius of horizon $r_{H}$, i.e. $\theta =0$ and $%
\theta =\pi $, we take $X\rightarrow 0$ and $P\rightarrow 1$ . Finally, on
the horizon, we initially take $X=0$ and $P=1.$

We employ a polar grid of points $(r_{i},\theta _{j}),$ where $r$ goes from $%
r_{H}$ to some large value of $r$ ( $r_{\infty }$) which is much greater
than $r_{H}$ and $\theta $ runs from $0$ to $\pi .$ \ We use the finite
difference method and rewrite the non linear partial differential equation (%
\ref{eqx}) as 
\begin{equation}
A_{ij}X_{i+1,j}+B_{ij}X_{i-1,j}+C_{ij}X_{i,j+1}+D_{ij}X_{i,j-1}+E_{ij}X_{i,j}=F_{ij}
\label{findiffxx}
\end{equation}

where $X_{ij}=X(r_{i},\theta _{j}).$ For the interior grid points, the
coefficients $A_{ij},...,F_{ij\text{ }}$ are given by, 
\begin{equation}
\begin{tabular}{l}
$A_{ij}=-\frac{1}{r_{i}\Delta r}(1-\frac{m}{r_{i}}+2\frac{r_{i}^{2}}{l^{2}})-%
\frac{1}{(\Delta r)^{2}}(1-\frac{2m}{r_{i}}+\frac{r_{i}^{2}}{l^{2}})$ \\ 
$B_{ij}=\frac{1}{r_{i}\Delta r}(1-\frac{m}{r_{i}}+2\frac{r_{i}^{2}}{l^{2}})-%
\frac{1}{(\Delta r)^{2}}(1-\frac{2m}{r_{i}}+\frac{r_{i}^{2}}{l^{2}})$ \\ 
$C_{ij}=-\frac{1}{2r_{i}^{2}\Delta \theta }\cot \theta _{j}-\frac{1}{
(r_{i}\Delta \theta )^{2}}$ \\ 
$D_{ij}=\frac{1}{2r_{i}^{2}\Delta \theta }\cot \theta _{j}-\frac{1}{
(r_{i}\Delta \theta )^{2}}$ \\ 
$E_{ij}=\frac{2}{(\Delta r)^{2}}(1-\frac{2m}{r_{i}}+\frac{r_{i}^{2}}{l^{2}})+%
\frac{2}{(r_{i}\Delta \theta )^{2}}+(\frac{NP_{ij}}{r_{i}\sin \theta })^{2}$
\\ 
$F_{ij}=-\frac{1}{2}X_{ij}(X_{ij}^{2}-1)$%
\end{tabular}
\label{xcoeff}
\end{equation}

and for the points on the horizon $(i=1)$, the coefficients are given by

\begin{equation}
\begin{tabular}{l}
$A_{1j}=B_{1j}=0$ \\ 
$C_{1j}=-\frac{1}{4\Delta \theta }\cot \theta _{j}-\frac{1}{2\Delta \theta
^{2}}$ \\ 
$D_{1j}=\frac{1}{4\Delta \theta }\cot \theta _{j}-\frac{1}{2\Delta \theta
^{2}}$ \\ 
$E_{1j}=\frac{1}{\Delta \theta ^{2}}+\frac{1}{2}(\frac{NP_{1j}}{\sin \theta }
)^{2}$ \\ 
$F_{1j}=\frac{X_{2j}-X_{1j}}{\Delta r}r_{H}(1-\frac{m}{r_{H}}+\frac{
2r_{H}^{2}}{l^{2}})-\frac{r_{H}^{2}}{4}(X_{1j}^{2}-1)$%
\end{tabular}
\label{xhcoeff}
\end{equation}

The equation (\ref{eqp}) could be rewritten as the same as the finite
difference equation (\ref{findiffxx}) by replacing $X_{ij}$ to $P_{ij}$ and
the following form for the coefficients inside the grid 
\begin{equation}
\begin{tabular}{l}
$A_{ij}^{\prime }=-\frac{1}{(\Delta r)^{2}}(1-\frac{2m}{r_{i}}+\frac{%
r_{i}^{2}}{l^{2}})-\frac{1}{\Delta r}(\frac{m}{r_{i}^{2}}+\frac{r}{l^{2}})$
\\ 
$B_{ij}^{\prime }=-\frac{1}{(\Delta r)^{2}}(1-\frac{2m}{r_{i}}+\frac{%
r_{i}^{2}}{l^{2}})+\frac{1}{\Delta r}(\frac{m}{r_{i}^{2}}+\frac{r}{l^{2}})$
\\ 
$C_{ij}^{\prime }=\frac{1}{2r_{i}^{2}\Delta \theta }\cot \theta _{j}-\frac{1%
}{(r_{i}\Delta \theta )^{2}}$ \\ 
$D_{ij}^{\prime }=-\frac{1}{2r_{i}^{2}\Delta \theta }\cot \theta _{j}-\frac{1%
}{(r_{i}\Delta \theta )^{2}}$ \\ 
$E_{ij}^{\prime }=\frac{2}{(\Delta r)^{2}}(1-\frac{2m}{r_{i}}+\frac{r_{i}^{2}%
}{l^{2}})+\frac{2}{(r_{i}\Delta \theta )^{2}}+2X_{ij}^{2}$ \\ 
$F_{ij}^{\prime }=0$%
\end{tabular}
\label{pcoeff}
\end{equation}

On the horizon, the coefficients of the finite difference equation of $P$
field are

\begin{equation}
\begin{tabular}{l}
$A_{1j}^{\prime }=B_{1j}^{\prime }=0$ \\ 
$C_{1j}^{\prime }=\frac{1}{4\Delta \theta }\cot \theta _{j}-\frac{1}{2\Delta
\theta ^{2}}$ \\ 
$D_{1j}^{\prime }=-\frac{1}{4\Delta \theta }\cot \theta _{j}-\frac{1}{
2\Delta \theta ^{2}}$ \\ 
$E_{1j}^{\prime }=\frac{1}{\Delta \theta ^{2}}+r_{H}^{2}X_{1j}^{2}$ \\ 
$F_{1j}^{\prime }=-\frac{P_{2j}-P_{1j}}{\Delta r}r_{H}(\frac{m}{r_{H}}+\frac{
r_{H}^{2}}{l^{2}})$%
\end{tabular}
\label{phcoeff}
\end{equation}

Now, by using the well known successive overrelaxation method \cite{num} for
the above mentioned finite difference equations, we obtain the values of $X$
and $P$ fields inside the grid, which we denote them by $X^{(1)}$ and $%
P^{(1)}$. Then by calculating the $r$-gradients of $X$ and $P$ just outside
the horizon and iterating the finite difference equations on the horizon, we
get the new values of $X$ and $P$ fields on the horizon points. Then these
new values of $X$ and $P$ fields are used as the new boundary condition on
the horizon for the next step in obtaining the values of $X$ and $P$ fields
inside the grid which could be denoted by $X^{(2)}$ and $P^{(2)}$. In the
successive overrelaxation method, the value of the each field in the $(n+1)$
-th iteration is related to the $n$-th iteration by

\begin{equation}
X_{ij}^{(n+1)}=X_{ij}^{(n)}-\omega \frac{\zeta _{ij}^{(n)}}{E_{ij}^{(n)}}
\label{sor}
\end{equation}
where residual matrix $\zeta _{ij}^{(n)}$ is the difference between the left
and right hand sides of the equation (\label{findiffx}\ref{findiffxx}),
evaluated in the $n$-th iteration and $\omega $ is the overrelaxation
parameter. The iteration is performed many times to some value $n=K,$ such
that $\sum_{i,j}\left| X_{ij}^{K}-X_{ij}^{K-1}\right| <\varepsilon $ for a
given error $\varepsilon $. It is a matter of trial and error to find the
value of $\omega $ that yields rapid convergence.

\begin{figure}[tbp]
\begin{center}
\epsfig{file=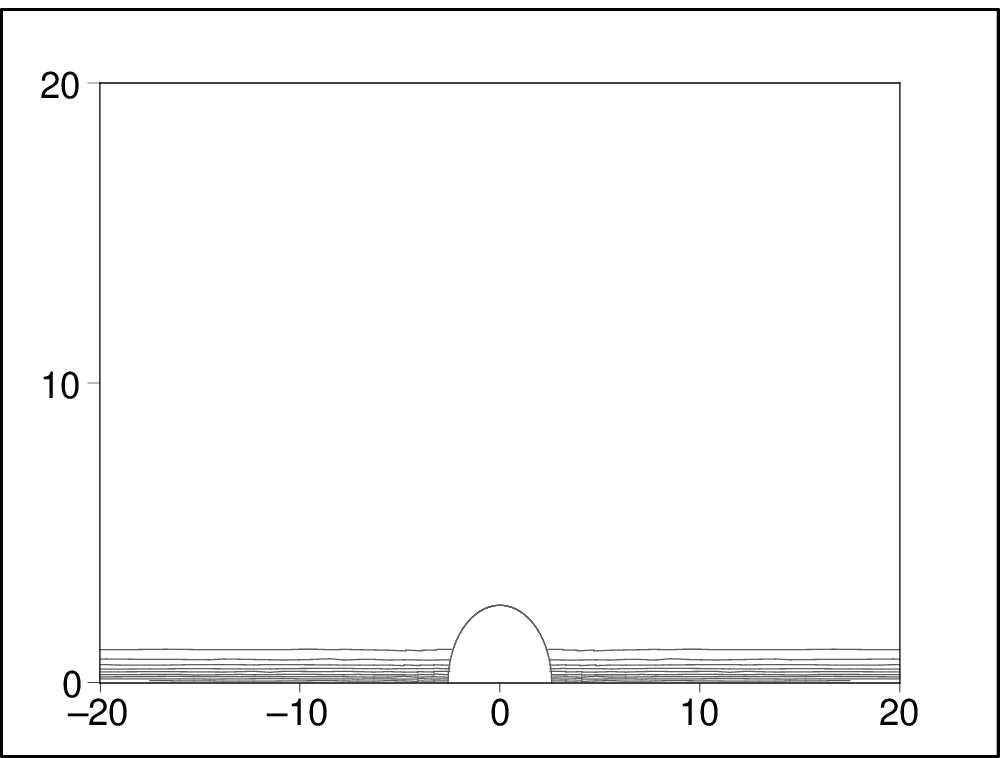,width=.4\linewidth} 
\epsfig{file=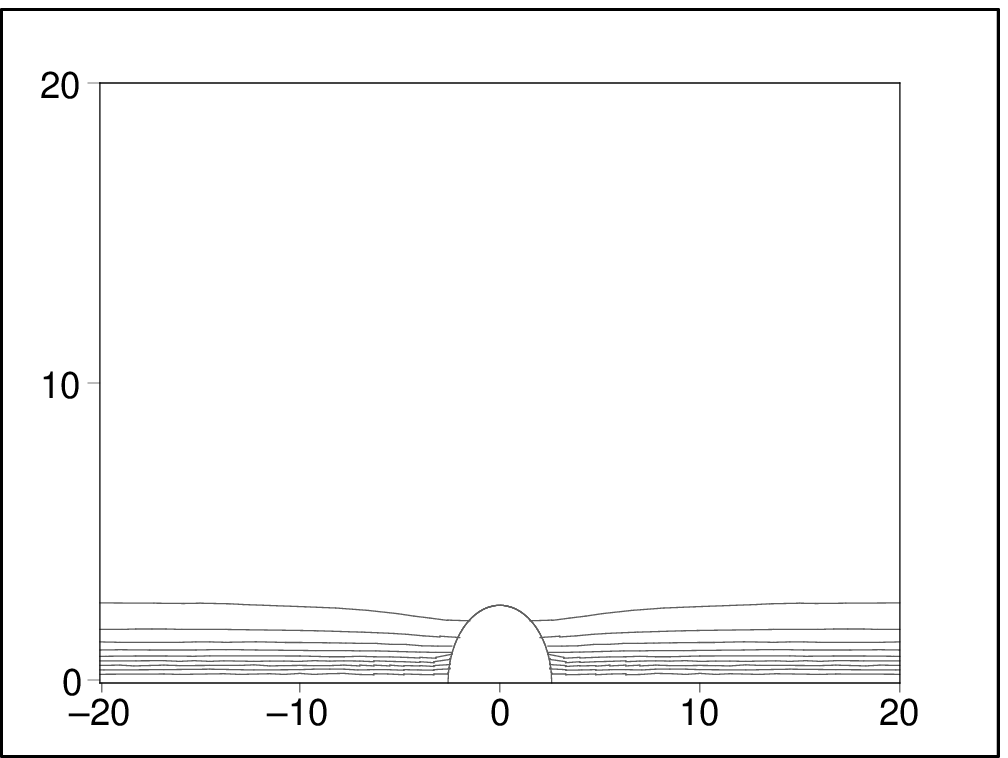,width=0.4
\linewidth}
\end{center}
\caption{{}$X$ and $P$ Contours ($X$ and $P$ increase from $0.1$ to $0.9$
upward and downward respectively) for $l=1$ and $N=1$.}
\label{fig4}
\end{figure}

\begin{figure}[tbp]
\begin{center}
\epsfig{file=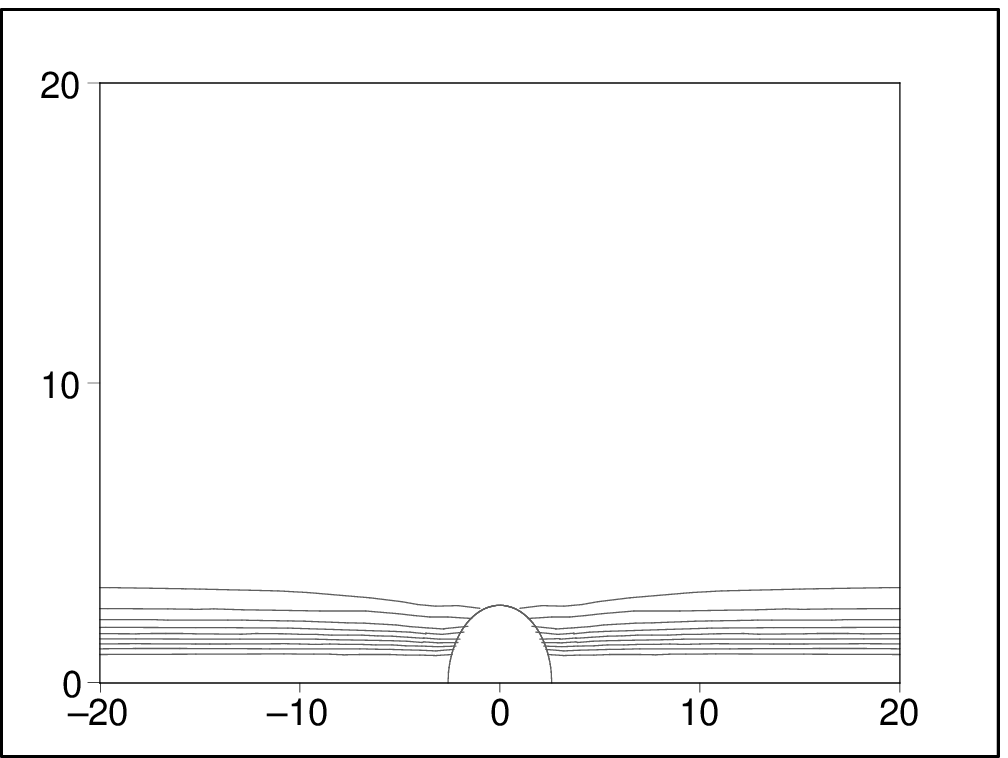,width=0.4\linewidth} %
\epsfig{file=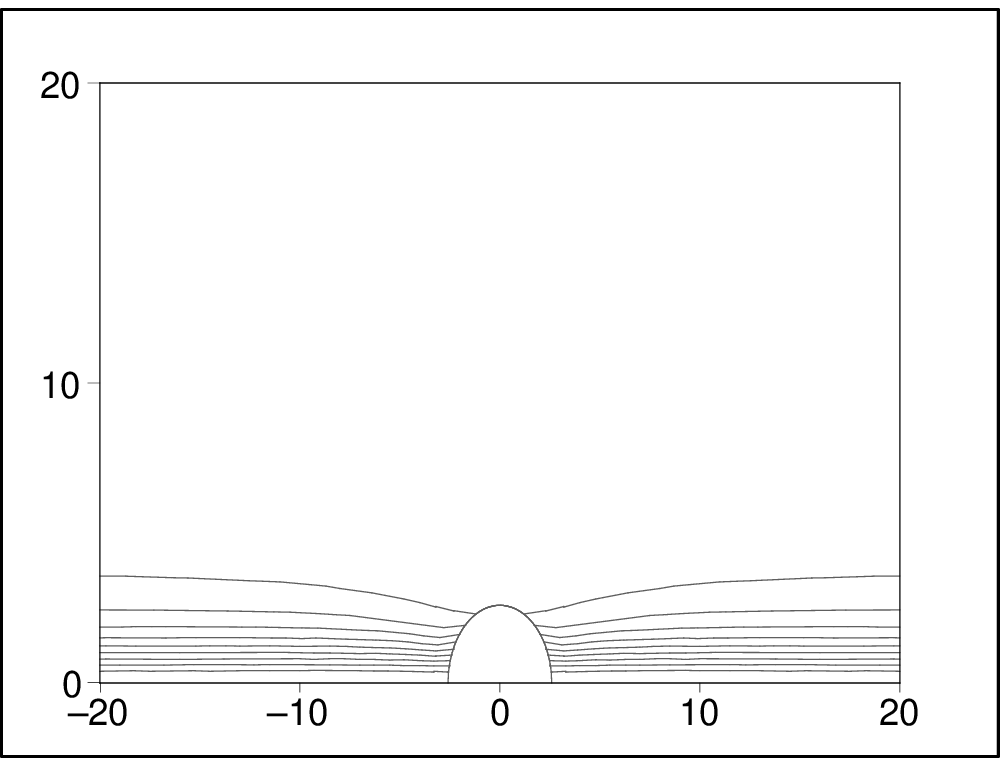,width=0.4\linewidth}
\end{center}
\caption{{}$X$ and $P$ Contours ($X$ and $P$ increase from $0.1$ to $0.9$
upward and downward respectively) for $l=1$ and $N=10$.}
\label{fig5}
\end{figure}

\begin{figure}[tbp]
\begin{center}
\epsfig{file=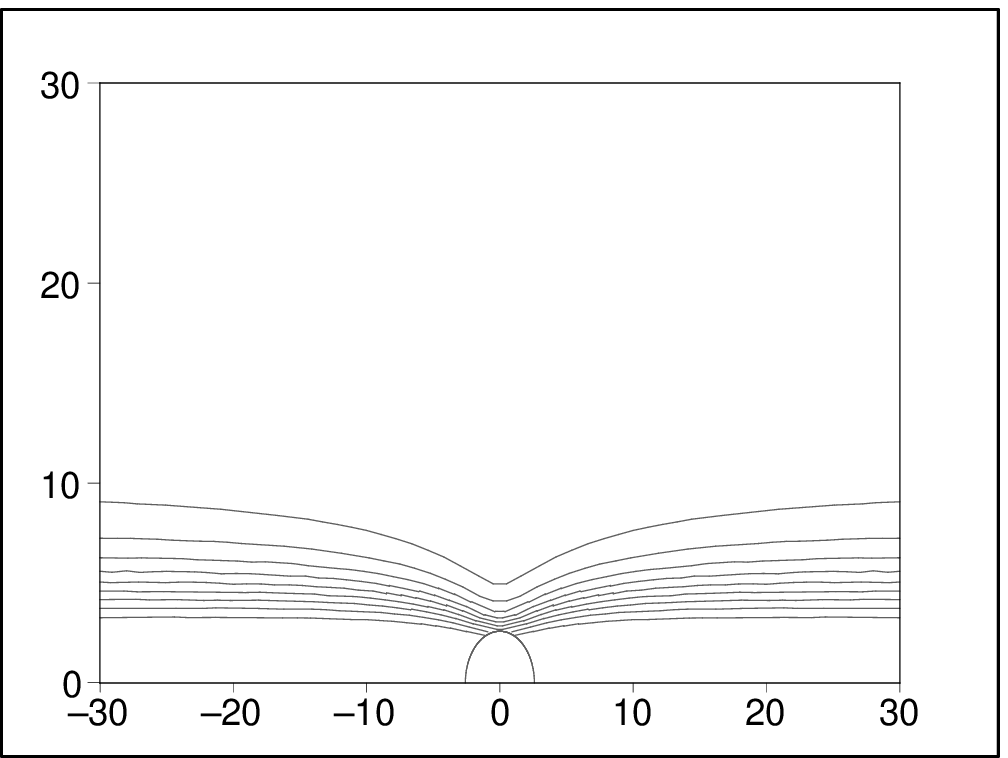,width=0.4\linewidth} %
\epsfig{file=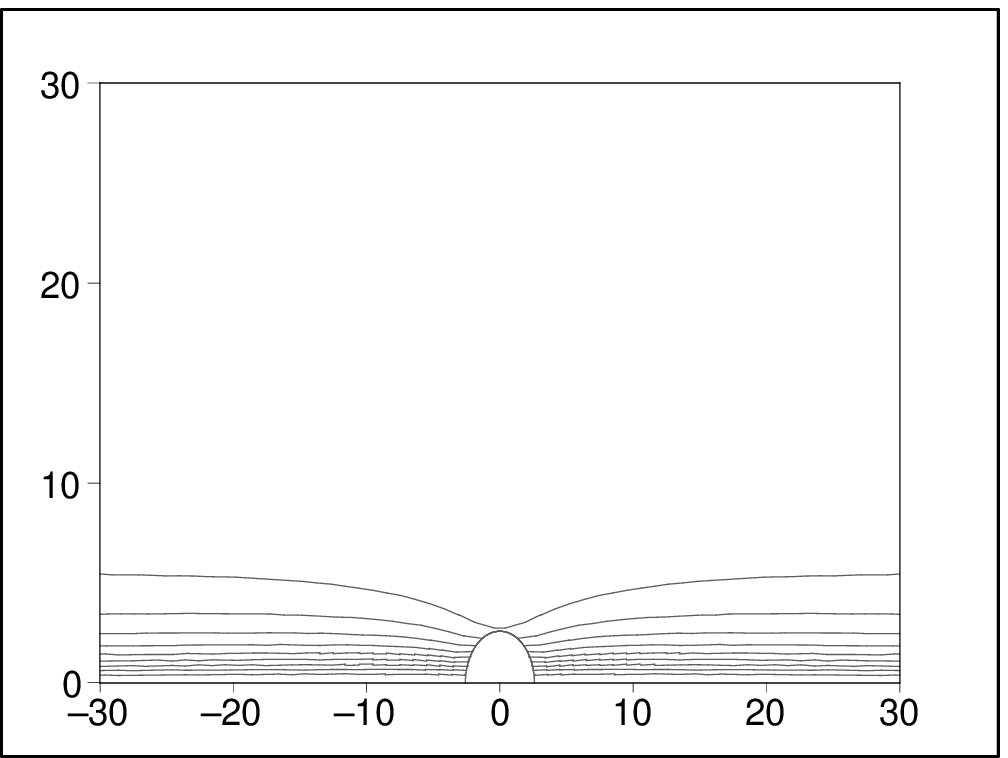,width=0.4\linewidth}
\end{center}
\caption{{}$X$ and $P$ Contours ($X$ and $P$ increase from $0.1$ to $0.9$
upward and downward respectively) for $l=1$ and $N=100$.}
\label{fig6}
\end{figure}

\begin{figure}[tbp]
\begin{center}
\epsfig{file=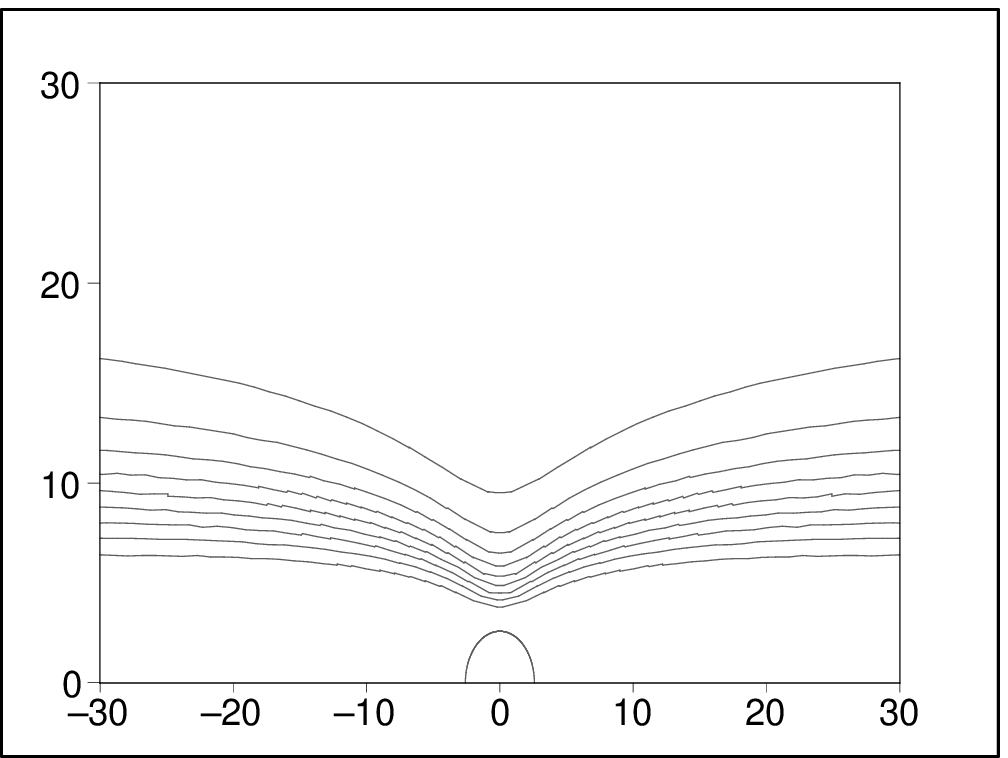,width=0.4\linewidth} %
\epsfig{file=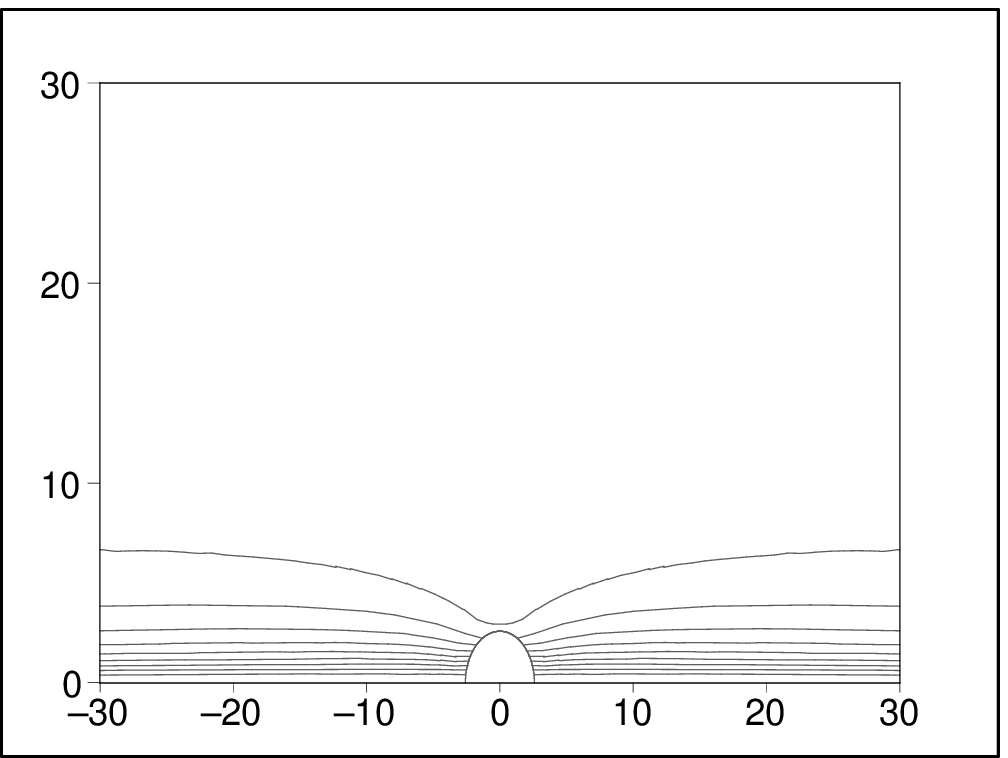,width=0.4\linewidth}
\end{center}
\caption{{}$X$ and $P$ Contours ($X$ and $P$ increase from $0.1$ to $0.9$
upward and downward respectively) for $l=1$ and $N=400$.}
\label{fig7}
\end{figure}

\begin{figure}[tbp]
\begin{center}
\epsfig{file=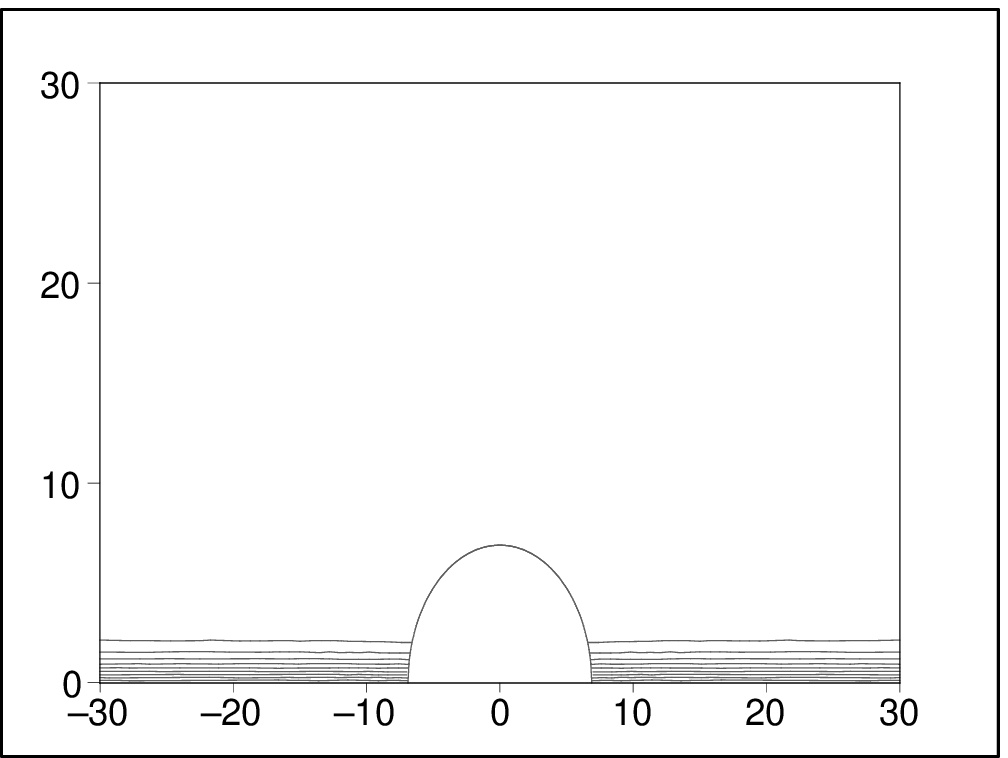,width=0.4\linewidth} 
\epsfig{file=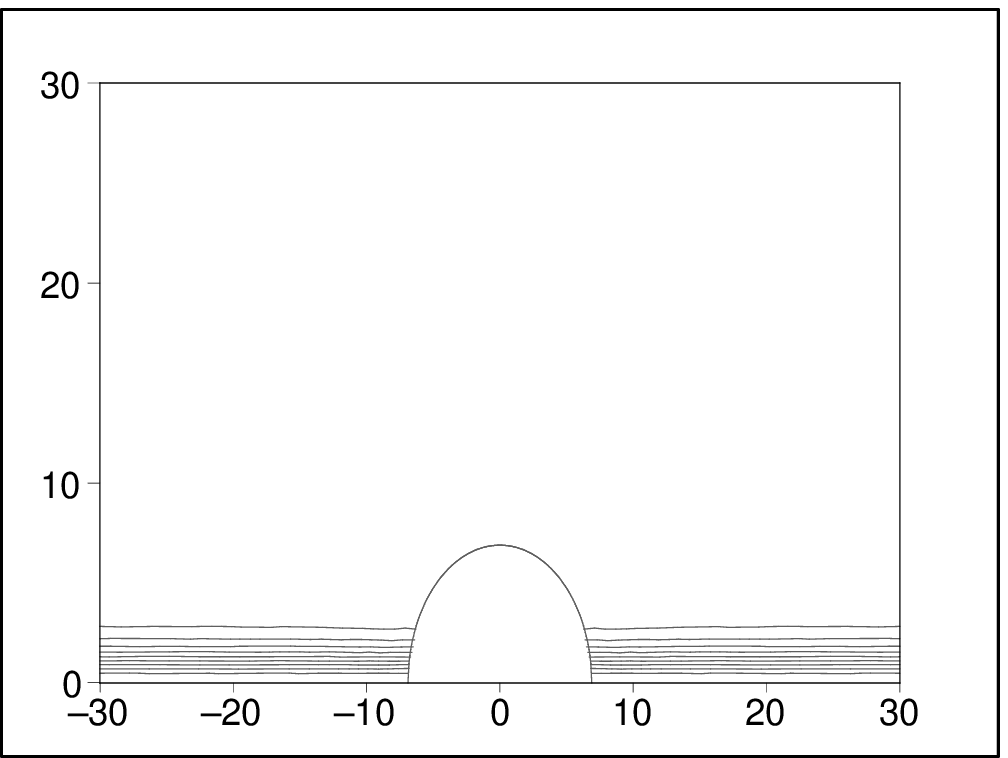,width=0.4
\linewidth}
\end{center}
\caption{{}$X$ and $P$ Contours ($X$ and $P$ increase from $0.1$ to $0.9$
upward and downward respectively) for $l=5$ and $N=1$.}
\label{fig8}
\end{figure}

\begin{figure}[tbp]
\begin{center}
\epsfig{file=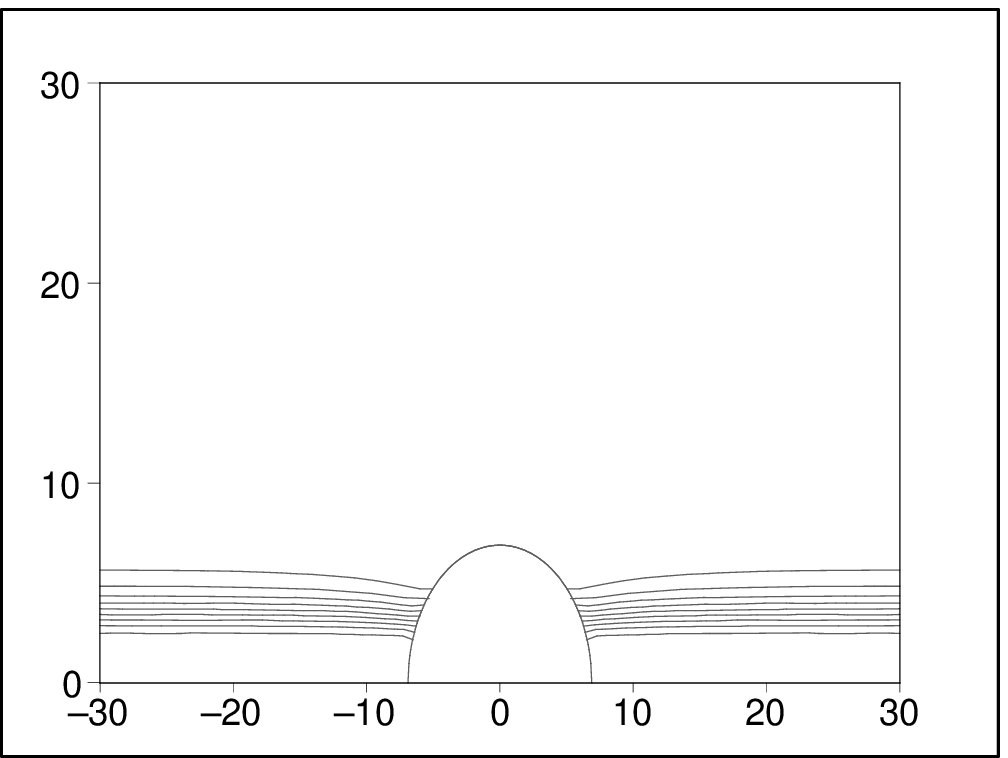,width=0.4\linewidth} %
\epsfig{file=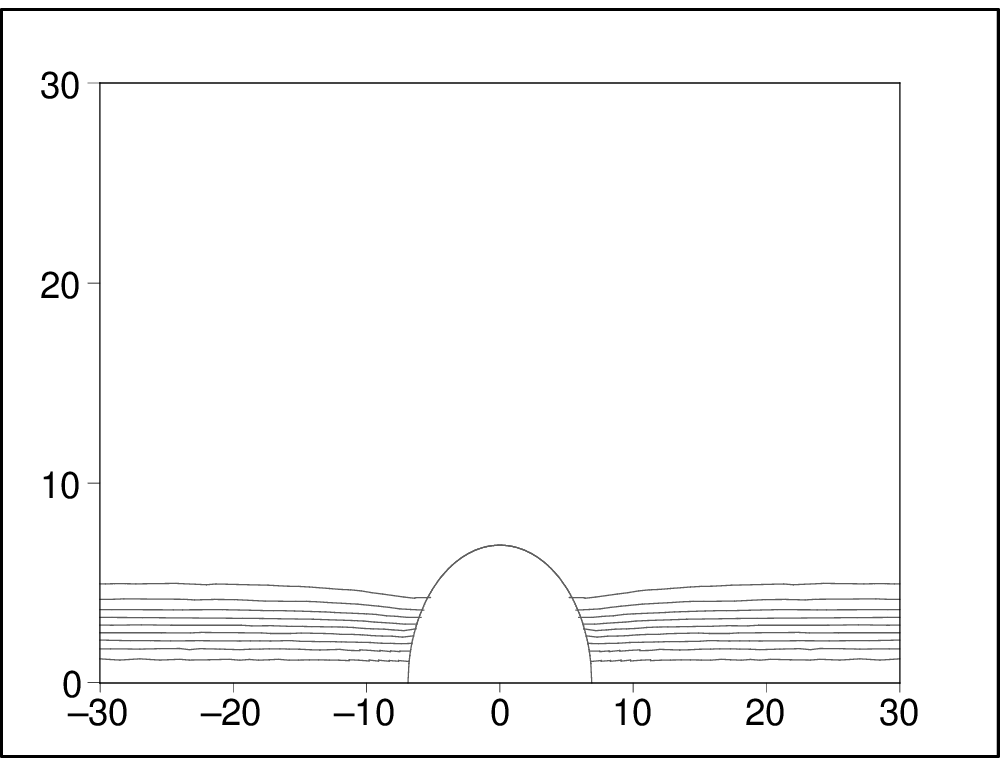,width=0.4\linewidth}
\end{center}
\caption{{}$X$ and $P$ Contours ($X$ and $P$ increase from $0.1$ to $0.9$
upward and downward respectively) for $l=5$ and $N=10$.}
\label{fig9}
\end{figure}

\begin{figure}[tbp]
\begin{center}
\epsfig{file=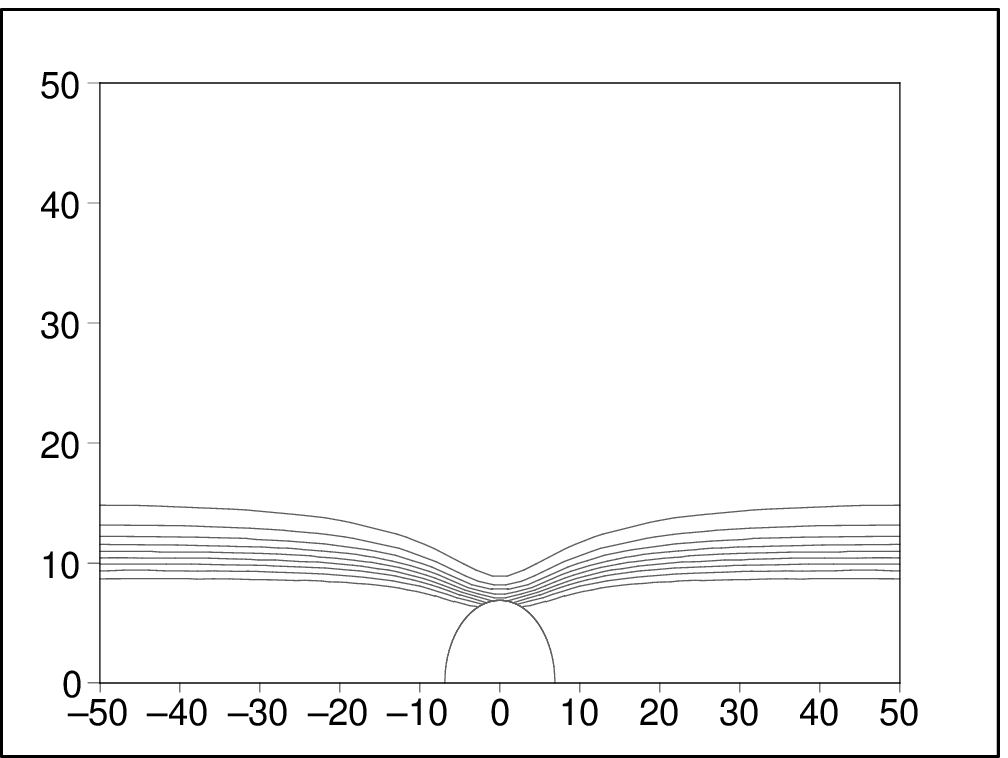,width=0.4\linewidth} %
\epsfig{file=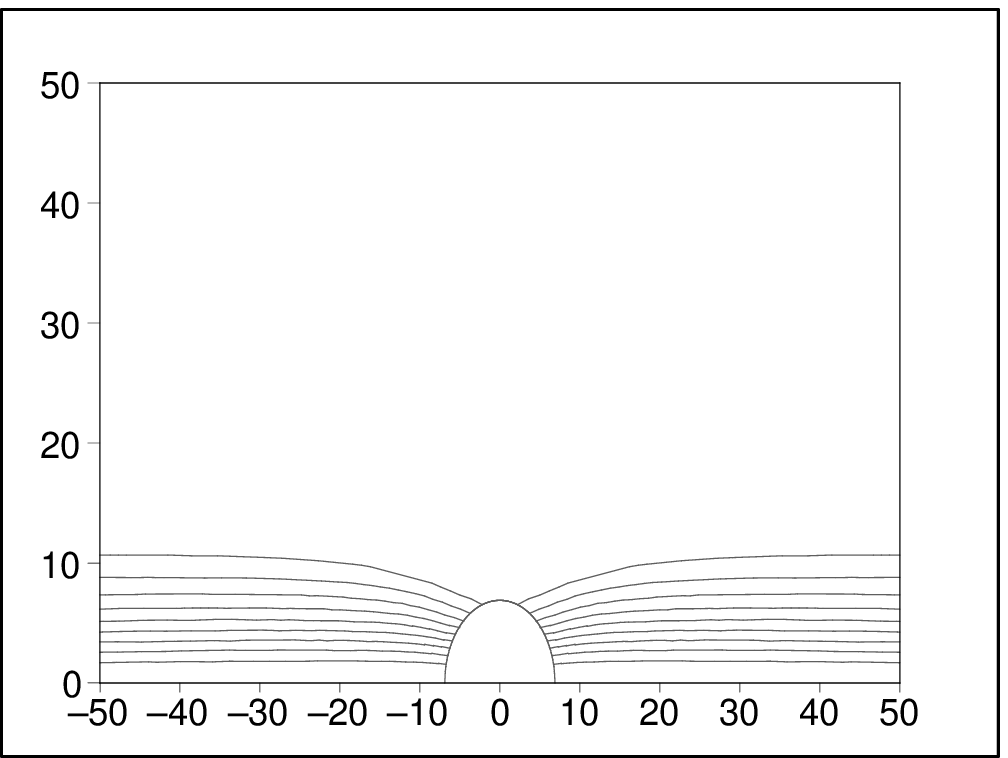,width=0.4\linewidth}
\end{center}
\caption{{}$X$ and $P$ Contours ($X$ and $P$ increase from $0.1$ to $0.9$
upward and downward respectively) for $l=5$ and $N=100$.}
\label{fig10}
\end{figure}

\begin{figure}[tbp]
\begin{center}
\epsfig{file=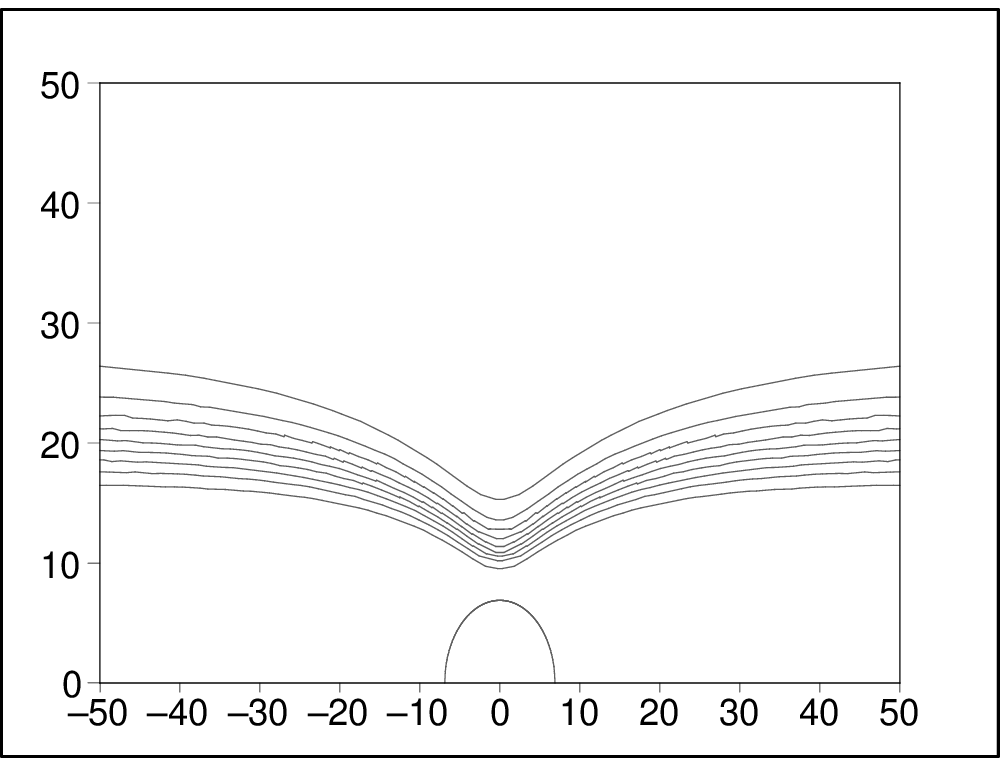,width=0.4\linewidth} %
\epsfig{file=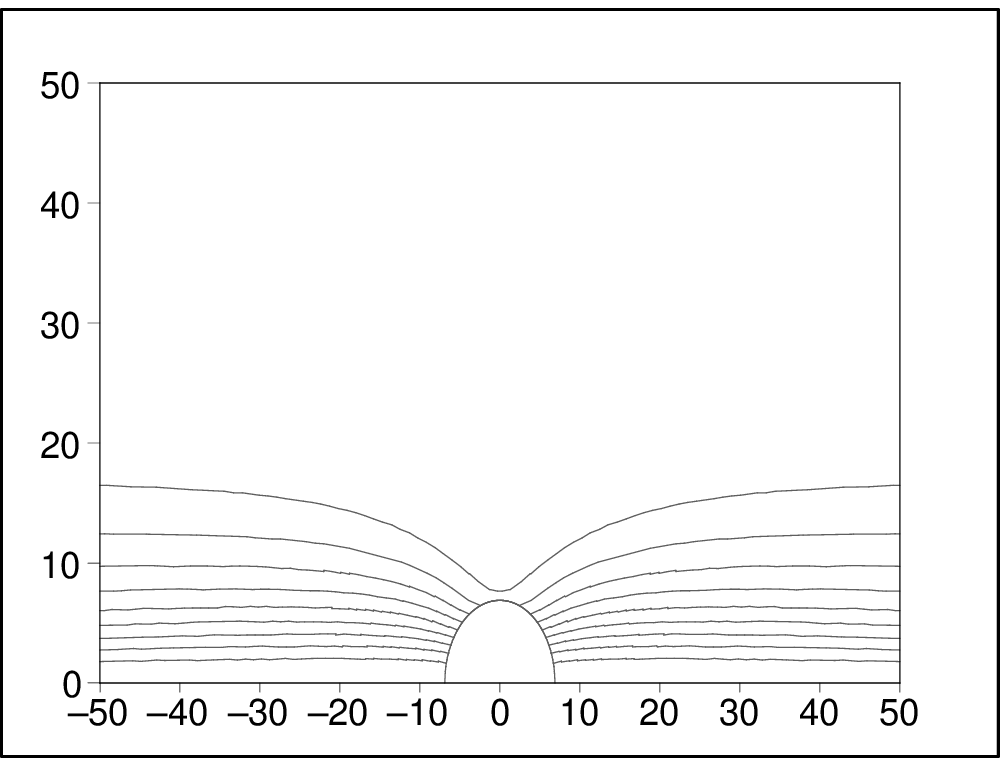,width=0.4\linewidth}
\end{center}
\caption{{}$X$ and $P$ Contours ($X$ and $P$ increase from $0.1$ to $0.9$
upward and downward respectively) for $l=5$ and $N=400$.}
\label{fig11}
\end{figure}

\begin{figure}[tbp]
\begin{center}
\epsfig{file=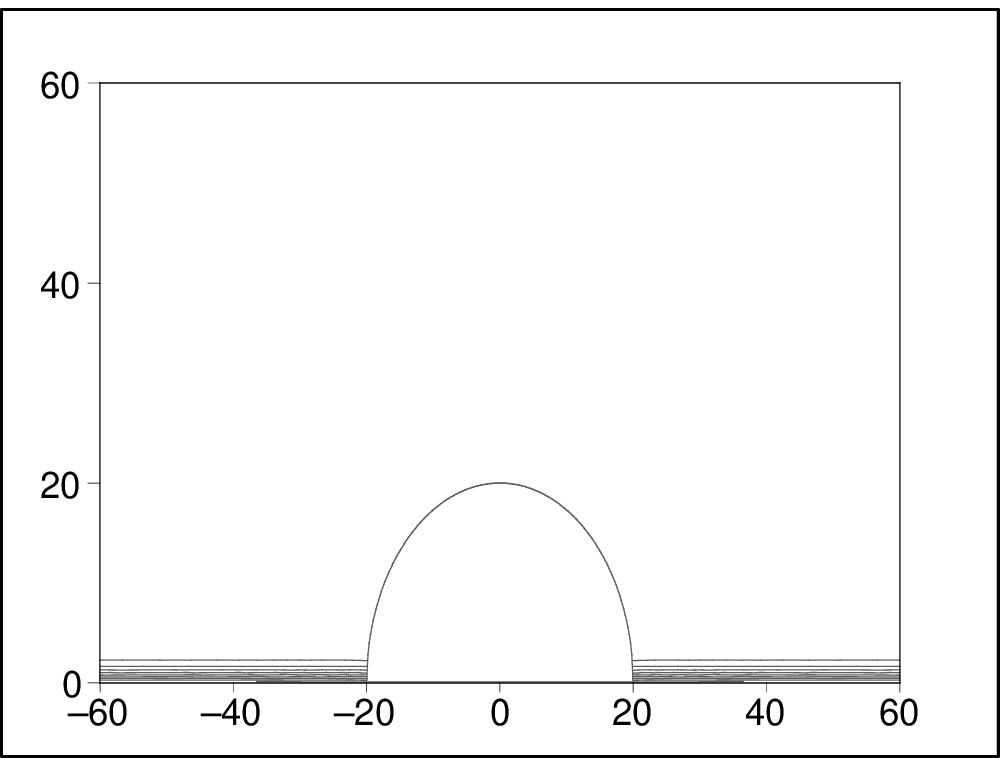,width=0.4\linewidth} %
\epsfig{file=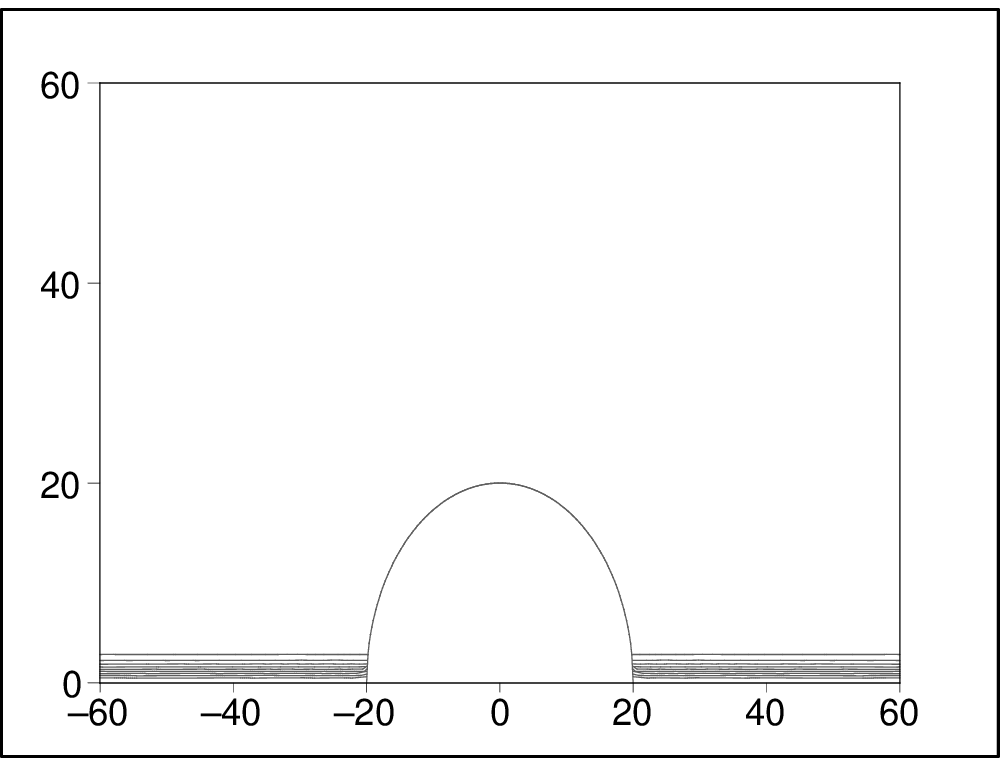,width=0.4\linewidth}
\end{center}
\caption{{}$X$ and $P$ Contours ($X$ and $P$ increase from $0.1$ to $0.9$
upward and downward respectively) for $l\rightarrow \infty $ and $N=1$.}
\label{fig12}
\end{figure}

\begin{figure}[tbp]
\begin{center}
\epsfig{file=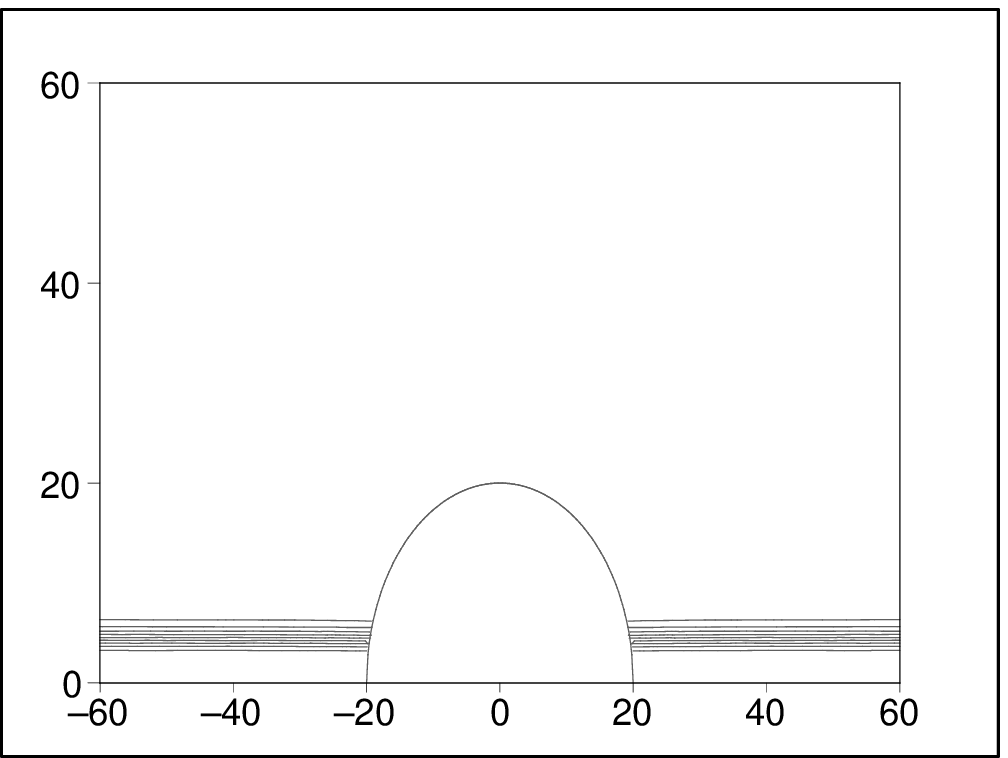,width=0.4\linewidth} %
\epsfig{file=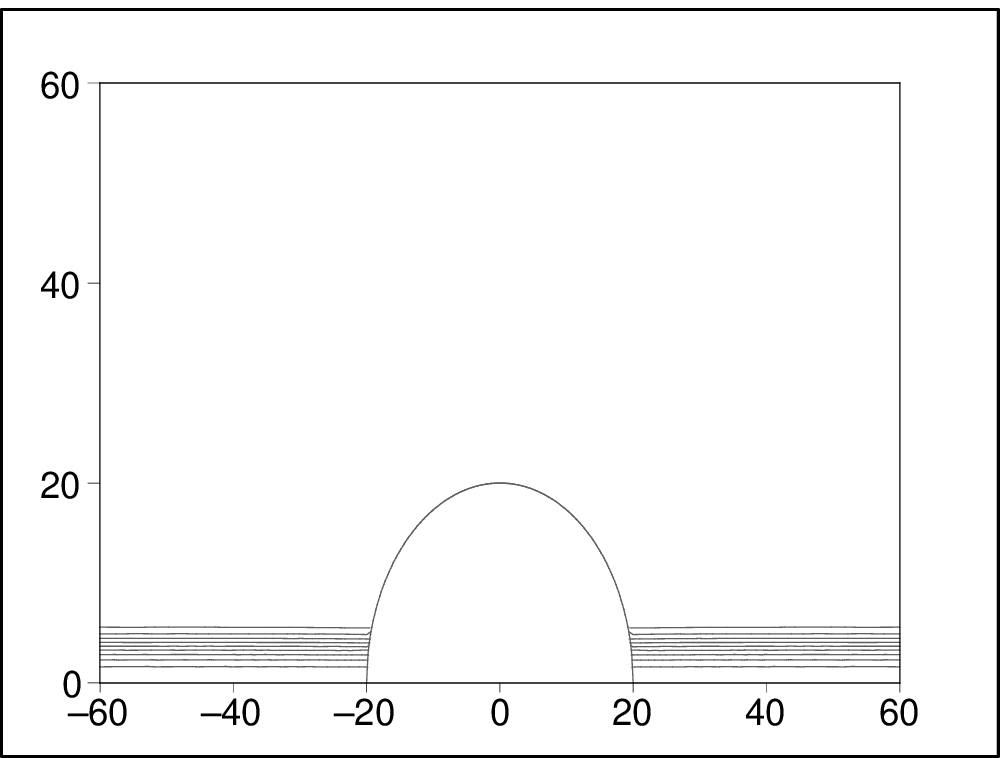,width=0.4\linewidth}
\end{center}
\caption{{}$X$ and $P$ Contours ($X$ and $P$ increase from $0.1$ to $0.9$
upward and downward respectively) for $l\rightarrow \infty $ and $N=10$.}
\label{fig13}
\end{figure}

\begin{figure}[tbp]
\begin{center}
\epsfig{file=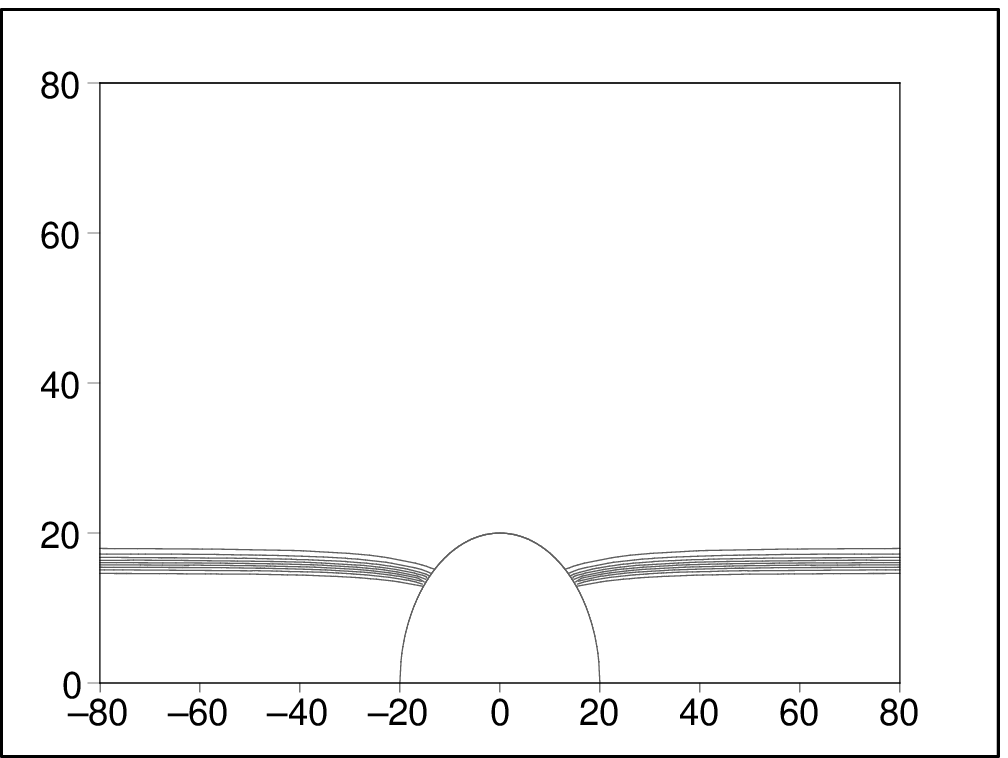,width=0.4\linewidth} %
\epsfig{file=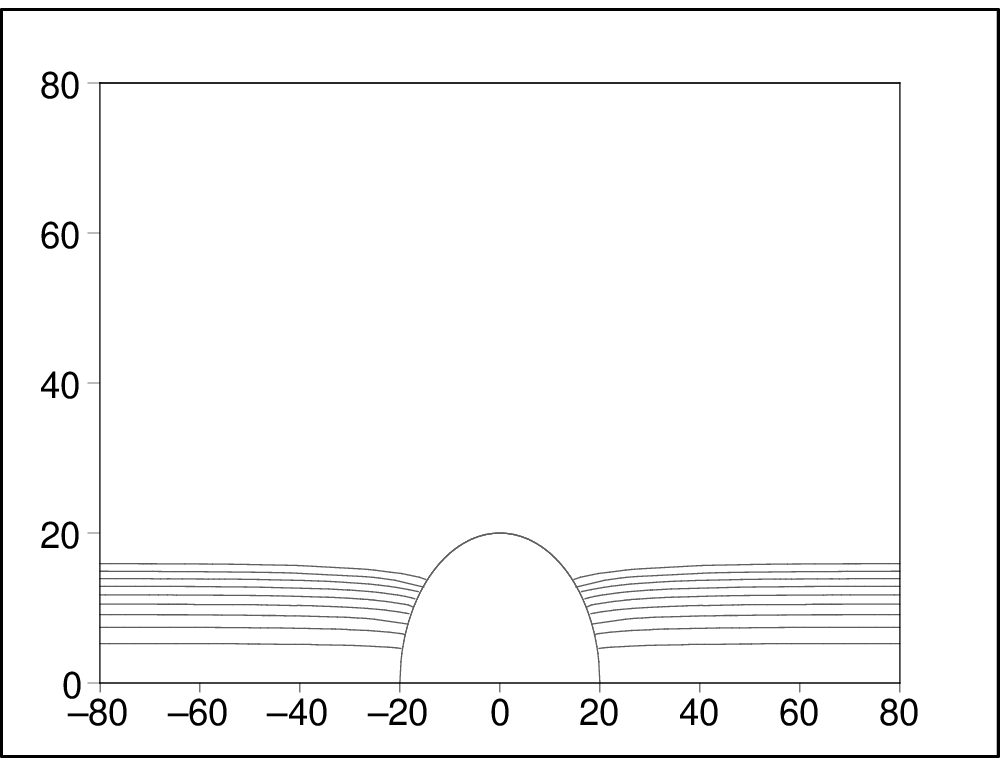,width=0.4\linewidth}
\end{center}
\caption{{}$X$ and $P$ Contours ($X$ and $P$ increase from $0.1$ to $0.9$
upward and downward respectively) for $l\rightarrow \infty $ and $N=100$.}
\label{fig14}
\end{figure}

\begin{figure}[tbp]
\begin{center}
\epsfig{file=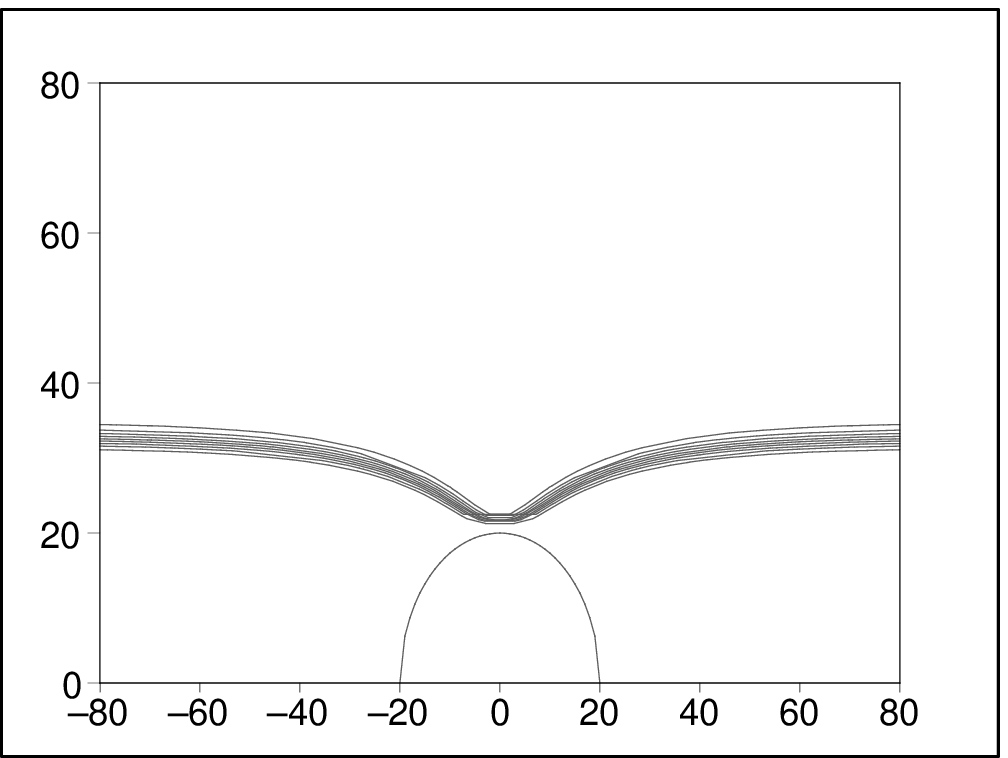,width=0.4\linewidth} %
\epsfig{file=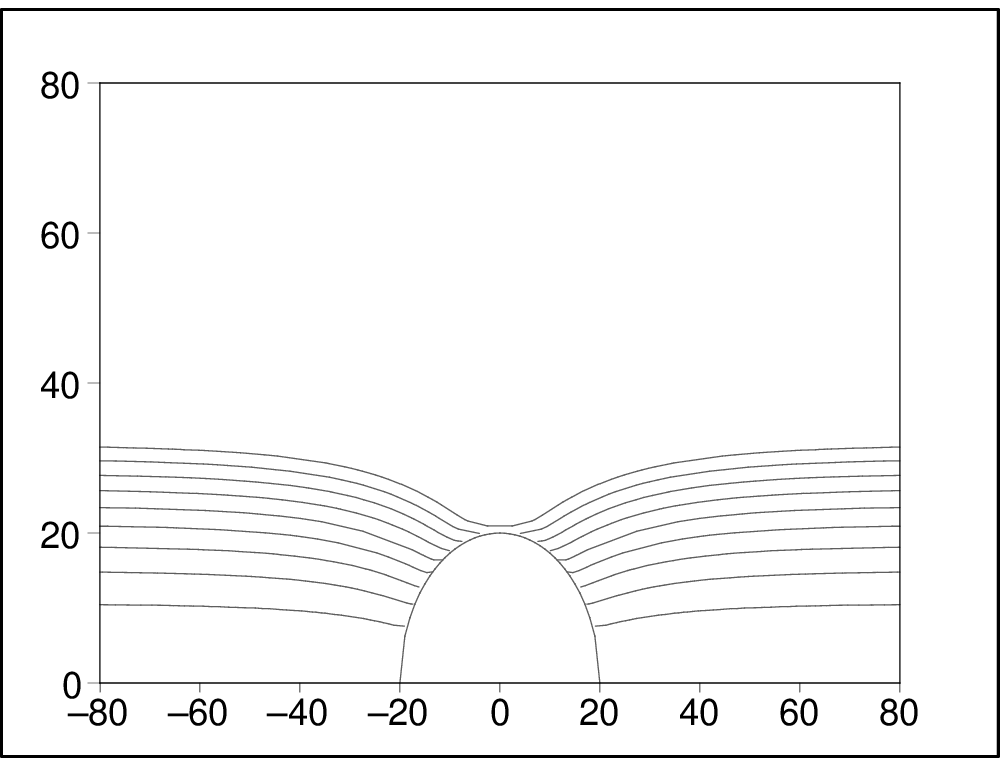,width=0.4\linewidth}
\end{center}
\caption{{}$X$ and $P$ Contours ($X$ and $P$ increase from $0.1$ to $0.9$
upward and downward respectively) for $l\rightarrow \infty $ and $N=400$.}
\label{fig15}
\end{figure}

The results of this calculation are displayed in figures (\ref{fig4}) to ( 
\ref{fig15}) for different values of $l=1,5$ and $l\rightarrow \infty $ \
and winding numbers $N=1,10,100,400.$ In all of these figures, the black
hole mass is taken have the constant value $m=10.$ Note that for $%
l\rightarrow \infty $ figures (\ref{fig12}-\ref{fig15}) are the same as the
figures introduced in \cite{Achu} for the Abelian-Higgs model in the flat
spacetime. Like the flat spacetime case, in the AdS spacetime, increasing
the winding number yields a greater vortex thickness. Comparing \ figures (%
\ref{fig6},\ref{fig10} and \ref{fig15}) we see that as $l$ decreases the
black hole is completely covered by a vortex of decreasingly large winding
number. For example, for $l=1$ the black hole horizon is completely inside
the core of a vortex with a winding numberof less than one hundred, but for $%
l\rightarrow \infty ,$ this occurs for winding number about four hundred.

\begin{figure}[tbp]
\begin{center}
\epsfig{file=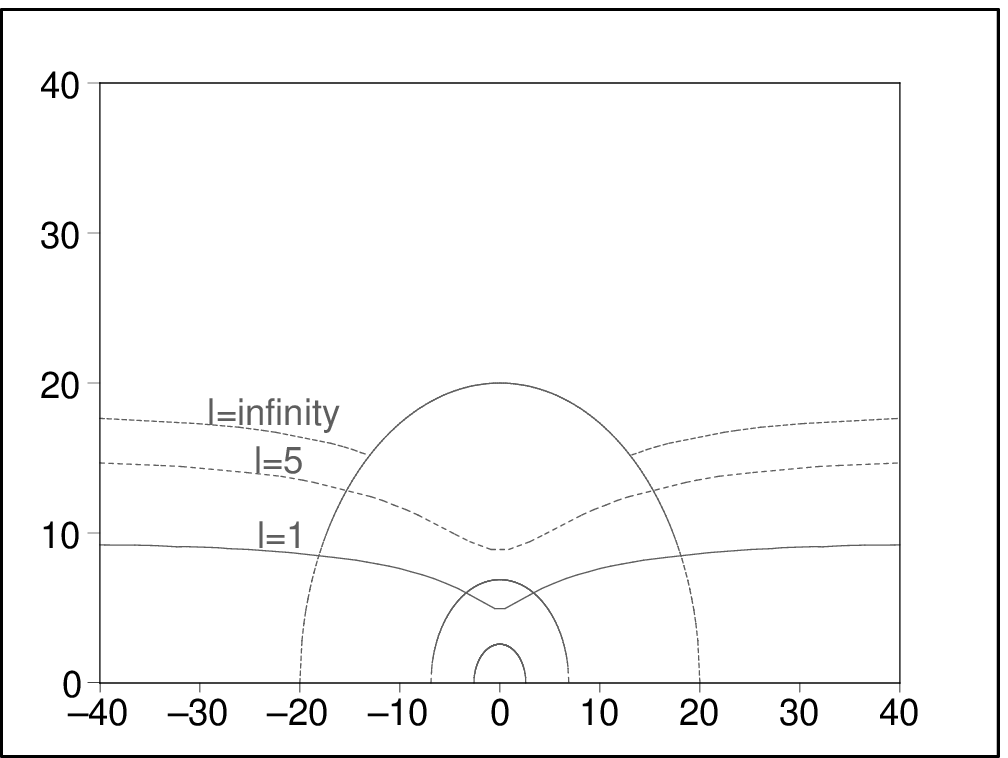,width=0.4\linewidth} %
\epsfig{file=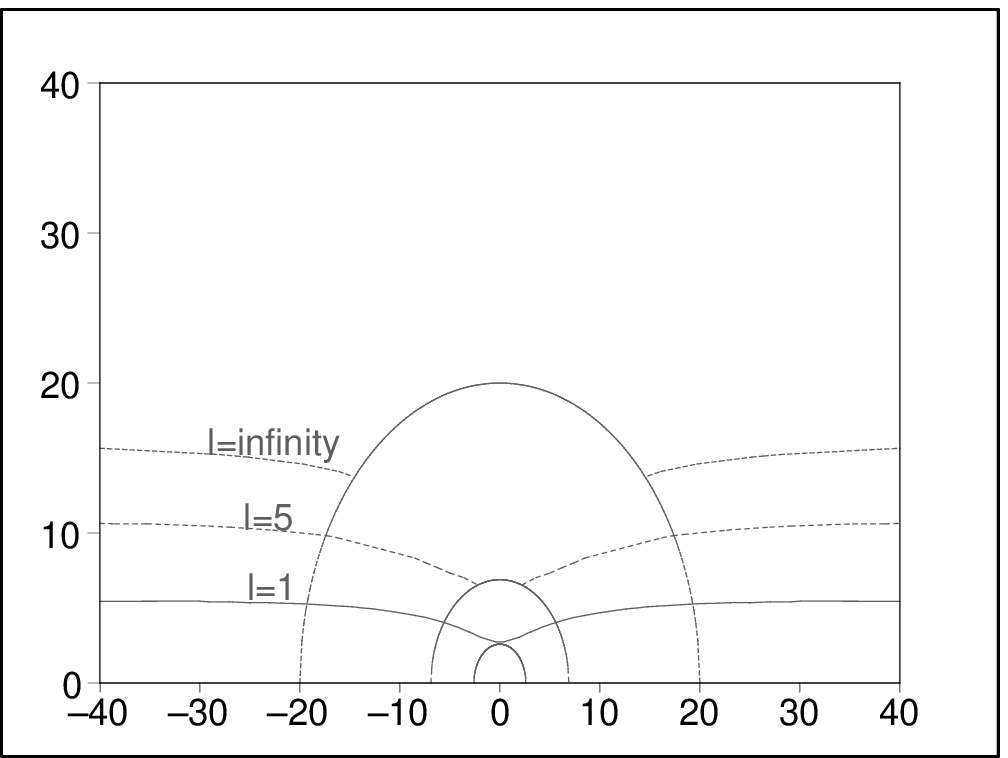,width=0.4\linewidth}
\end{center}
\caption{{}$X=0.9$ and $P=0.1$ Contours for $l=1$ (solid)$,5$ (dotted) and $%
l\rightarrow \infty $ (dashed) with winding number $N=100$.}
\label{fig16}
\end{figure}

\begin{figure}[tbp]
\begin{center}
\epsfig{file=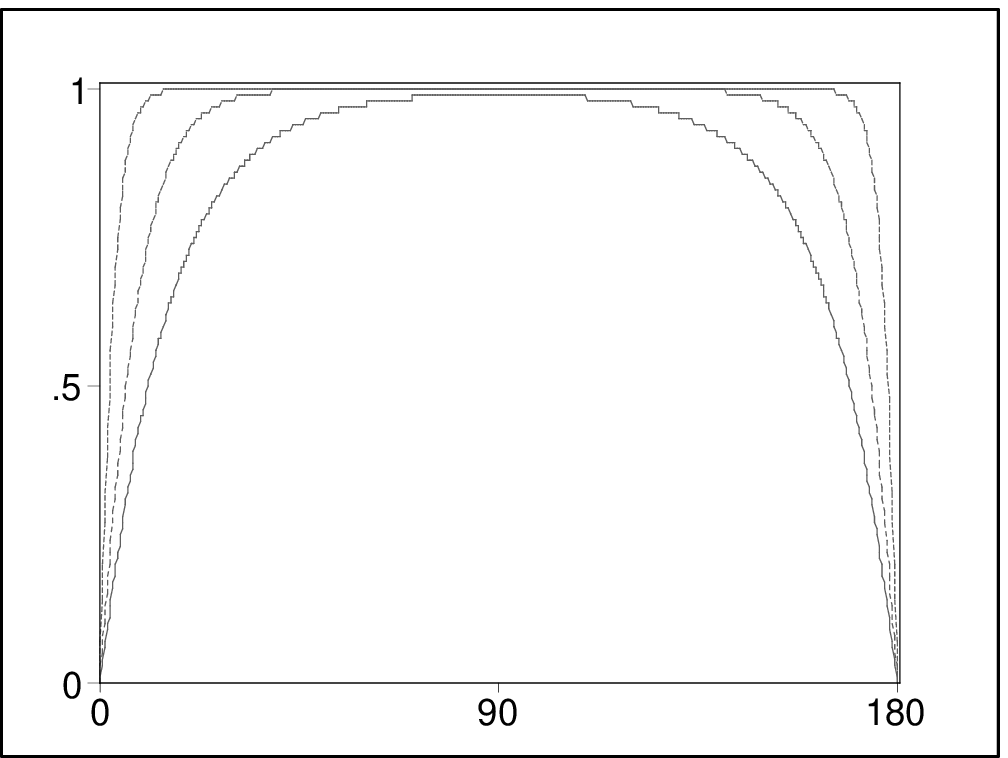,width=0.4\linewidth} %
\epsfig{file=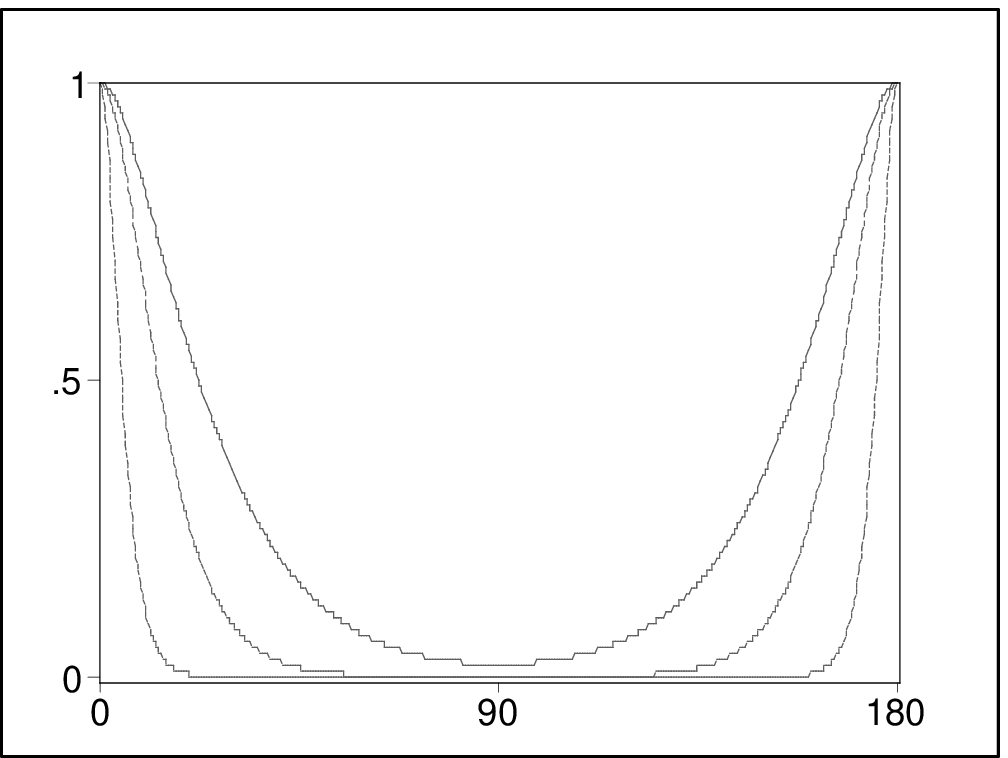,width=0.4\linewidth}
\end{center}
\caption{{}$X$ and $P$ fields on the horizon versus $\protect\theta $ for $%
l=1$ (solid)$,5$ (dotted) and $l\rightarrow \infty $ (dashed) with winding
number $N=1.$ }
\label{fig17}
\end{figure}

Also, as figure (\ref{fig16}) shows, for a string with definite winding
number, the string core increases with increasing $l,$ but the ratio of
string core to the black hole horizon decreases. Alternatively from figure (%
\ref{fig17}) we see the $X$ \ and $P$ fields more rapidly approach their
respective maximum and minimum values in smaller angle as $l$ increases.

\section{Vortex Self Gravity on the AdS-Schwarzschild Black hole}

We now consider the effect of \ the vortex on the AdS-Schwarzschild black
hole. As we have seen in section (2), this is a formidable problem even for
flat or AdS spacetimes.

As in section (2), we assume that the thickness of string is much smaller
that all the other relevant length scales and the gravitational effects of
the string are weak enough so that the linearized Einstein-Abelian Higgs
differential equations are applicable. So, we condsider thin string \ with
the winding number $N=1$ in the AdS-Schwarzschild background with $l=1$. The
rescaled components of the energy-momentum tensor are 
\begin{equation}
\begin{tabular}{l}
$T_{tt}=(1-\frac{2m}{r}+\frac{r^{2}}{l^{2}})\{\frac{1}{2}(1-\frac{2m}{r}+%
\frac{r^{2}}{l^{2}})[(\frac{\partial X}{\partial r})^{2}+\frac{1}{\alpha
r^{2}\sin ^{2}\theta }(\frac{\partial P}{\partial r})^{2}]+\frac{1}{2r^{2}}(%
\frac{\partial X}{\partial \theta })^{2}+\frac{1}{2\alpha r^{4}\sin
^{2}\theta }(\frac{\partial P}{\partial \theta })^{2}$ \\ 
$\ \ \ \ +(X^{2}-1)^{2}+\frac{1}{2r^{2}\sin ^{2}\theta }X^{2}P^{2}\}$ \\ 
\\ 
$T_{rr}=\frac{1}{2}(\frac{\partial X}{\partial r})^{2}-\frac{1}{2r^{2}}(1-%
\frac{2m}{r}+\frac{r^{2}}{l^{2}})^{-1}[(\frac{\partial X}{\partial \theta }%
)^{2}+\frac{1}{\alpha r^{2}\sin ^{2}\theta }(\frac{\partial P}{\partial
\theta })^{2}]+\frac{1}{2\alpha r^{2}\sin ^{2}\theta }(\frac{\partial P}{%
\partial r})^{2}$ \\ 
$\ \ \ \ -(1-\frac{2m}{r}+\frac{r^{2}}{l^{2}})^{-1}[(X^{2}-1)^{2}+\frac{1}{%
2r^{2}\sin ^{2}\theta }X^{2}P^{2}]$ \\ 
\\ 
$T_{\theta \theta }=-\frac{1}{2}(1-\frac{2m}{r}+\frac{r^{2}}{l^{2}})[r^{2}(%
\frac{\partial X}{\partial r})^{2}+\frac{1}{\alpha \sin ^{2}\theta }(\frac{%
\partial P}{\partial r})^{2}]+\frac{1}{2}(\frac{\partial X}{\partial \theta }%
)^{2}+\frac{1}{2\alpha r^{2}\sin ^{2}\theta }(\frac{\partial P}{\partial
\theta })^{2}$ \\ 
$\ \ \ \ -r^{2}(X^{2}-1)^{2}-\frac{1}{2\sin ^{2}\theta }X^{2}P^{2}$ \\ 
\\ 
$T_{\varphi \varphi }=-\frac{1}{2}(1-\frac{2m}{r}+\frac{r^{2}}{l^{2}}%
)[r^{2}\sin ^{2}\theta (\frac{\partial X}{\partial r})^{2}-\frac{1}{\alpha }(%
\frac{\partial P}{\partial r})^{2}]-\frac{1}{2}\sin ^{2}\theta (\frac{%
\partial X}{\partial \theta })^{2}+\frac{1}{2\alpha r^{2}}(\frac{\partial P}{%
\partial \theta })^{2}$ \\ 
$\ \ \ \ \ -r^{2}\sin ^{2}\theta (X^{2}-1)^{2}+\frac{1}{2}X^{2}P^{2}$%
\end{tabular}
\label{stress}
\end{equation}

\begin{figure}[tbp]
\begin{center}
\epsfig{file=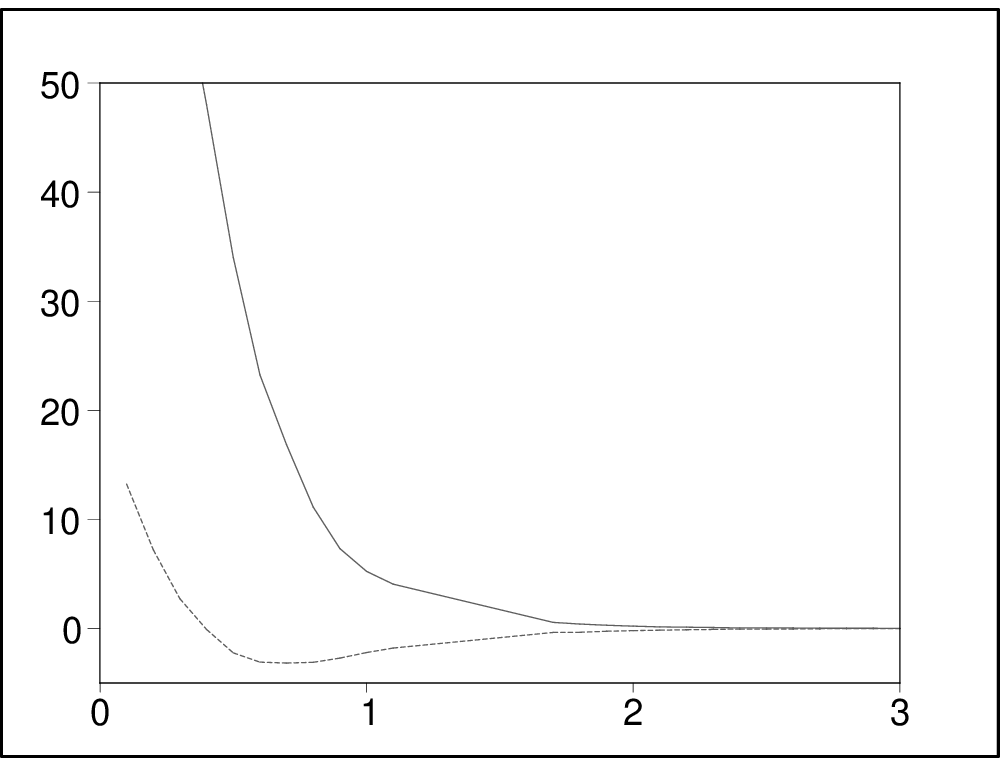,width=0.4\linewidth} %
\epsfig{file=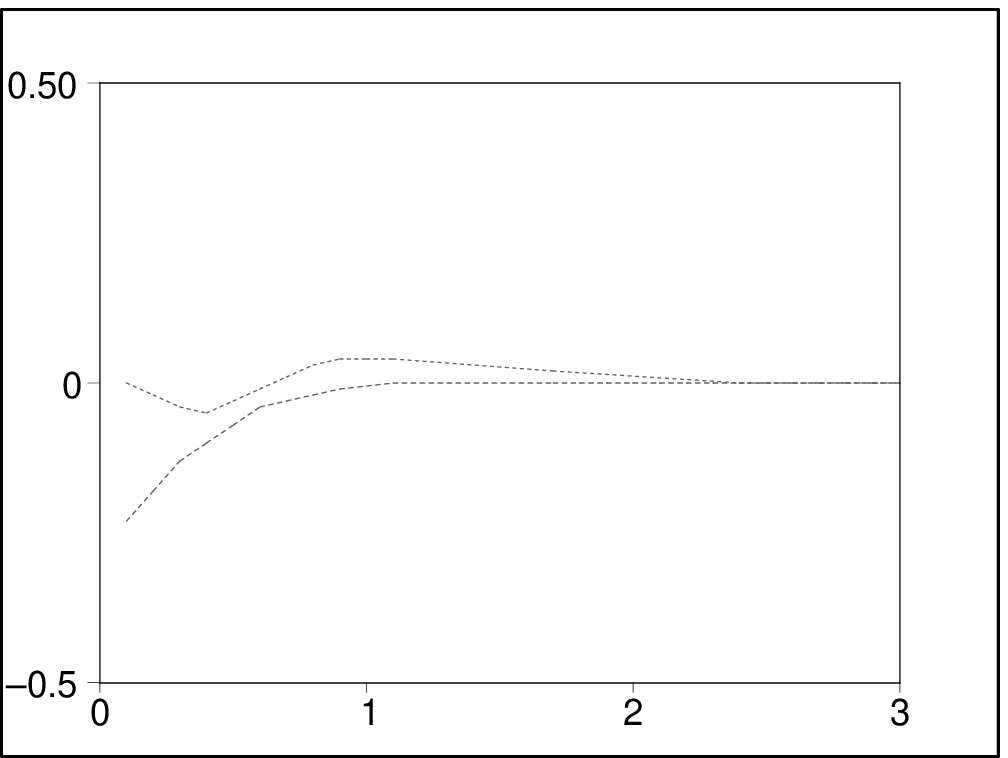,width=0.4\linewidth}
\end{center}
\caption{{}$T_{tt}$ (solid), $T_{\protect\theta \protect\theta }$ (dashed), $%
T_{\protect\varphi \protect\varphi }$ (dotted) and $T_{rr}$ (dot-dashed)
curves versus $\protect\rho $ in $\ z=5$ for the AdS-Schwarzschild black
hole with $l=1$ and winding number $N=1.$ }
\label{fig18}
\end{figure}

In the figures (\ref{fig18}), the behaviour of the energy-momentum tensor
components in a fixed $z$ is shown. The behaviour of the components in the
other $z$ is like the same as these figures. As it is clear from figures,
the components of the energy-momentunm tensor rapidly go to zero outside the
core string, so the situation is like what happened in the pure AdS
spacetime. Then performing the same calculation as in the pure AdS spacetime
described in detail in section (2), give us the following metric of the
AdS-Schwarzschild spacetime incorporated the effect of the vortex on it,

\begin{equation}
ds^{2}=-(1-\frac{2m}{r}+\frac{r^{2}}{l^{2}})dt^{2}+\frac{dr^{2}}{1-\frac{2m}{%
r}+\frac{r^{2}}{l^{2}}}+r^{2}(d\theta ^{2}+\beta ^{2}\sin ^{2}\theta d\phi
^{2})  \label{ADSSCHDEF}
\end{equation}

The above metric describes an AdS-Schwarzschild metric with a deficit angle.
So, using a physical Lagrangian based model, we have established that the
presence of the cosmic string induces a deficit angle in the black hole
metric.

\section{Polystring Configuration}

From the precceding sections we know that an AdS-Schwarzschild \ black hole
could support a long range Nielsen-Olesen vortex string as stable hair. The
natural next question is, {\it how many strings could be supported by such a
black hole?}

Such a question was considered for the Schwarzschild black hole. {\it \ }In 
\cite{Dow}, a three string configuration in Schwarzscild black hole was
studied. Also, in \cite{Frolov}, all possible multistring configurations
were presented.

Since the intersection of strings with the AdS-Schwarzschild black hole
occurs on some points on the horizon which is topologically $S^{2},$ and
such a situation also holds for the Schwarzschild black hole whose horizon
has spherical topology, we could use the same procedure as was presented in 
\cite{Frolov}. So, we give an argument in brief to show the different
possible multistring-AdS Schwarzschild black hole system.

We assume strings enter the horizon along radii. The different polystring
configurations are obtained by demanding that the black hole-multistring
system must be invariant under any rotation about any axis which coincides
with the strings. This condition is necessary since the black
hole-multistring system must be in force-free equilibrium. The intersection
of strings with the black hole occur on some points on the horizon which is
topologically $S^{2},$ and these points are the vertexes of a spherical
tesselation. The spherical tesselation is obtained by projection of the
edges of a polyhedron from its geometrical center onto a concentric sphere.
Every edge and vertex of the polyhedron is mapped to an arc of a great
circle on the sphere and a vertex of spherical tesselation respectively. The
spherical tesselation is invariant under a discrete rotation group. The
elements of this discrete group are just rotation around every axis which
passes through the center of the sphere and vertices, mid-arc points and
centers of \ faces of \ spherical tesselation, respectively. This discrete
rotation group of the spherical tesselation is in correspondence with the
rotational symmetry group of the polyhedron. So the black hole-polystring
system with radial strings is invariant under discrete rotations and hence
is in an equilibrium state.

By knowing the properties of spherical tesselation, in \cite{Frolov}, the
different multistring configurations are introduced. The first configuration
is 14 strings which are pierced to the black hole on the symmetry axis of a
tetrahedron. The second and third configurations are 26 and 62 strings which
are pierced to the black hole on the symmetry axis of a octahedron and
icosahedron respectively. The fourth configuration is 2N strings (N is any
arbitrary number) which corresponds to the symmetry axis of a double pyramid.

\bigskip

\section{Conclusion}

The effect of a vortex on pure AdS spacetime to create a deficit angle in
the metric in the thin vortex approximation. We have extended this result,
establishing numerically that Abelian Higgs vortices of finite thickness can
pierce an AdS-Schwarzschild black hole horizon. These solutions could thus
be interpreted as stable Abelian hair for the black hole. We have obtained
numerical solutions for various cosmological constants and string winding
numbers. Our solutions in the limit of \ $l\rightarrow \infty ,$ coincides
with the known solutions in the asymptotically flat spacetime.

We found that by increasing the winding number, the string core increases.
Also, the generalization to include piercing of more strings to the black
hole has been considered and it is shown that there are four different
polystring configurations. Finally, inclusion of the self gravity of the
vortex in the AdS-Schwarzschild background metric was shown to induce a
deficit angle in the AdS-Schwarzschild metric.

Other related problems such as study of the vortex in the charged or
rotating black hole backgrounds and non-abelian vortex solution in
asymtotically AdS spacetime remain to be carried out. Another issue is the
holographic description of vortex solution in these asymptotically AdS
spacetimes. Work on these problems is in progress.

\bigskip

{\Large Acknowledgments}

This work was supported by the Natural Sciences and Engineering Research
Council of Canada.

\pagebreak 

\end{document}